\documentclass[11pt]{article}
\usepackage[margin=0.9in]{geometry}

\usepackage{siunitx}
\usepackage{booktabs}
\usepackage{colortbl}
\usepackage{tikz}
\usepackage{float}
\usepackage{placeins}
\usepackage{graphicx}
\usepackage{subfig}
\usepackage{multirow}
\usepackage{amsmath,mathtools}
\usepackage{amsfonts}
\usepackage{soul}
\usepackage{comment}
\usepackage{enumitem}
\usepackage{mathrsfs}
\usepackage{hyperref}

\usepackage{nomencl}
\usepackage{tcolorbox}
\makenomenclature
\newtcolorbox[auto counter, number within=section]{nomenclaturebox}[2][]{colframe=black!10!white, colback=black!5!white, coltitle=black, fonttitle=\bfseries, title={}}
\usepackage{algorithm}
\usepackage[algo2e,ruled,vlined]{algorithm2e}
\usepackage{setspace}
\usepackage{cases}
\usepackage{authblk}
\usepackage{amsthm}
\usepackage[superscript]{cite}

\hypersetup{pdfauthor={Nojus Plungė}}

\newtcolorbox{highlighted}{colback=yellow,coltext=red,breakable}

\newcommand{\FIG}[1]{Fig.~\ref{#1}}

\newcommand{\EQ}[1]{Eq.~\eqref{#1}}

\newcommand{\TAB}[1]{Table~\ref{#1}}

\newlength\myindent
\newcommand\BibTeX{{\rmfamily B\kern-.05em \textsc{i\kern-.025em b}\kern-.08em
T\kern-.1667em\lower.7ex\hbox{E}\kern-.125emX}}

\begin{document}

\title{Deep learning-based phase-field modelling of brittle fracture in anisotropic media}

\author[1]{Nojus Plungė}
\author[1]{Peter Brommer}
\author[2]{Rachel S. Edwards}
\author[1]{Emmanouil G. Kakouris\thanks{Corresponding author: \texttt{Emmanouil.Kakouris@warwick.ac.uk}, School of Engineering, University of Warwick, Coventry, CV4 7AL, UK.}}

\affil[1]{School of Engineering, University of Warwick, UK}
\affil[2]{Department of Physics, University of Warwick, UK}
\date{}

\maketitle

\begin{abstract}{This work presents a variational physics-informed deep learning framework for phase-field modelling of brittle crack propagation in anisotropic media. Previous Deep Ritz Method (DRM) approaches have focused on second-order, isotropic phase-field fracture formulations and have relied on fully connected neural networks (NNs). In contrast, the present work introduces, for the first time within a variational deep learning setting, a family of higher-order anisotropic phase-field models through a generalised crack density functional. The fracture problem is solved by minimising the total energy using the DRM. The trial space is enriched with higher-order B-spline basis functions to represent higher-order gradients accurately and stably, thereby eliminating the need for conventional automatic differentiation. A ReZero-based residual NN architecture is adopted, for the first time in fracture mechanics. It stabilises the optimisation of the non-convex energy functional and improves convergence under strong coupling between displacement and phase fields. Random Fourier Feature (RFF) enrichment is also introduced, enhancing the representation of steep displacement gradients and narrow damage zones near crack tips. The methodology is assessed for isotropic behaviour and cubic and orthotropic anisotropies. Numerical examples demonstrate direction-dependent crack growth in anisotropic cases, highlighting the capability of the method to accurately capture this behaviour.}
\end{abstract}

\section{Introduction} \label{sec:intro}

Predictive modelling and simulation of fracture phenomena remain of central importance in engineering design, safety assessment, and structural integrity analysis. Fracture processes are inherently non-linear and strongly path-dependent, and are often characterised by complex crack topologies involving initiation, branching, deflection, and coalescence.\cite{Sharon1995, Sharon1996, Lee2011} Traditional fracture mechanics approaches, which rely on sharp crack representations and explicit tracking of crack surfaces, face substantial difficulties in accommodating such complexity, particularly when crack paths are not known a priori.\cite{Anderson2017} These limitations have motivated the development of more flexible and unified modelling frameworks.

Over the past two decades, the phase-field approach to fracture has emerged as a powerful alternative.\cite{Bourdin2008} Within this framework, cracks are represented implicitly through a continuous auxiliary field, commonly referred to as the phase field or damage variable. This field varies smoothly between values corresponding to intact and fully fractured material. The crack surface is therefore regularised over a finite width controlled by an internal length scale. The formulation is variational in nature and is typically derived from energy minimisation principles rooted in Griffith’s theory of brittle fracture.\cite{Griffith1921} Once discretised, the phase-field framework enables the simulation of complex fracture processes without the need for explicit crack tracking or ad hoc propagation criteria.\cite{Miehe2010,Borden2012} These advantages have led to widespread adoption of the method for a broad range of applications, including thermally induced cracking,\cite{Mandal2021} ductile fracture,\cite{Han2022} desiccation-induced cracking,\cite{Luo2023} and hydraulic fracture.\cite{Fei2023} The method has also demonstrated excellent capability in capturing three-dimensional crack patterns with quantitative accuracy.\cite{Borden2016}

In phase-field formulations, the mechanical response is described by two coupled fields: the displacement field and the phase field. The latter may be interpreted as a measure of material integrity, with values close to unity indicating intact material and values approaching zero corresponding to fully developed cracks. The governing equations follow from the stationarity conditions of a total energy functional that includes contributions from elastic strain energy, fracture surface energy, and external work. Despite its conceptual elegance, the phase-field approach poses two major computational challenges. Firstly, the governing energy functional is non-convex, which can lead to multiple local minima and demands robust numerical strategies to ensure convergence to physically meaningful solutions.\cite{Svolos2023} Secondly, accurate resolution of the internal length scale requires fine spatial discretisation, which substantially increases computational cost.\cite{Sarmadi2023} These challenges limit the applicability of conventional finite element implementations in large-scale simulations, optimisation, inverse problems, and uncertainty quantification.

Recent advances in machine learning have opened new avenues for the solution of partial differential equations and variational problems through physics-informed neural networks (PINNs).\cite{Raissi2019} In PINNs, NNs approximate the solution fields and are trained by minimising a loss function that penalises violations of the governing equations and boundary conditions. PINNs have been successfully applied to a wide range of problems in computational mechanics.\cite{Haghighat2021,Rao2021} However, when dealing with non-convex variational formulations, residual-based PINNs may suffer from optimisation difficulties and poor convergence.\cite{Basir2022}.

An alternative class of physics-informed methods is provided by variational or energy-based NN approaches. \cite{Goswami2020a, Goswami2020b, Samaniego2020} Among these, the Deep Ritz Method (DRM) has received particular attention.\cite{Yu2018} In the DRM, the NN represents the trial solution, and training is performed by directly minimising the energy functional of the problem. This approach bypasses the need to evaluate strong-form residuals and naturally aligns with variational formulations. The DRM has been shown to provide stable and accurate solutions for a variety of elliptic and variational problems.\cite{Yu2018} Its suitability for non-convex energies makes it especially attractive for phase-field fracture modelling.

Initial investigations of DRM-based solvers for phase-field fracture have demonstrated encouraging results for isotropic brittle fracture.\cite{Manav2024} These studies showed that NNs can learn crack nucleation and propagation by minimising the phase-field energy functional, without requiring labelled training data. Nevertheless, existing works have only focused on second-order isotropic phase-field formulations. Extensions to anisotropic fracture, and in particular to the higher-order crack density functionals required to model strong anisotropy, remain unexplored within variational deep learning frameworks.

Anisotropy plays a fundamental role in fracture processes in many materials, including geomaterials, composites, biological tissues, and crystalline solids.\cite{Ting1996,Wang2024,Bao1992} Direction-dependent fracture toughness leads to preferred crack paths and, in some cases, forbidden propagation directions.\cite{Gurtin1998,Takei2013} Phase-field formulations capable of capturing anisotropic surface energies have been developed by incorporating higher-order gradient terms and anisotropic tensors within the crack density functional.\cite{Li2015,Kakouris2018} These formulations enable the modelling of both weak and strong anisotropy, including cases in which the reciprocal surface energy becomes non-convex. However, conventional numerical implementations of such models are computationally demanding due to the presence of higher-order derivatives and stringent discretisation requirements.

The combination of anisotropic phase-field fracture models with variational deep learning therefore presents a compelling opportunity. On the one hand, anisotropic phase-field formulations provide a rigorous and flexible description of direction-dependent fracture. On the other hand, DRM-based neural solvers offer a mesh-free approximation space and naturally exploit the variational structure of the governing equations. Nevertheless, two key difficulties must be addressed. Firstly, higher-order anisotropic models involve fourth-order spatial derivatives of the phase field, which are challenging to approximate accurately using standard automatic differentiation alone. Secondly, the steep gradients of both displacement and phase fields near crack localisation demand trial spaces with sufficient smoothness and approximation power.

This work addresses these challenges by developing a variational physics-informed deep learning framework for anisotropic phase-field fracture. A family of higher-order anisotropic phase-field models is considered through a generalised crack density functional. The total energy functional is minimised using the DRM. To enable stable and accurate representation of higher-order gradients, the NN trial space is enriched using higher-order B-spline basis functions. This avoids numerical instabilities associated with evaluating higher-order derivatives via automatic differentiation in strongly non-convex settings. The boundary condition ansatz is generalised to accommodate arbitrary loading and constraint configurations within a unified expression. RFF enrichment \cite{Tancik2020} is introduced, for the first time in phase-field fracture problems, to improve the representation of steep spatial variations associated with crack localisation. A ReZero architecture \cite{Bachlechner2020} with residual connections (ResNet) \cite{He2015} is adopted, also for the first time in this work, to stabilise optimisation of the non-convex energy functional.

The proposed framework is assessed for isotropic, cubic, and orthotropic fracture surface energy densities. Numerical examples demonstrate direction-dependent crack growth in anisotropic cases, highlighting the capability of the method to accurately capture anisotropic fracture behaviour. The framework is additionally shown to capture crack kinking, a phenomenon arising from non-convexity of the energy landscape that induces abrupt changes in crack trajectory and necessitates convergence to a new energy minimum. Parametric studies examining the influence of the applied displacement discretisation and mesh resolution on the solution are also presented for the first time in this work. The results establish variational deep learning as a viable and flexible computational tool for anisotropic phase-field fracture.

The remainder of this paper is organised as follows. Section \ref{sec:phase_field_anisotr} summarises the governing equations of the anisotropic phase-field fracture model. Section \ref{sec:ext_deep_ritz_method} presents the DRM formulation and the proposed NN architecture. Section \ref{sec:num_results} reports numerical examples. Section \ref{sec:conclusions} concludes the paper.

\section{Phase field modelling for anisotropic fracture} \label{sec:phase_field_anisotr}

This section summarises the governing equations of the phase-field formulation for anisotropic brittle fracture, which form the theoretical basis of the deep learning framework developed in the subsequent sections.

\subsection{Variational description of fracture} \label{sec:variat_desc_frac}

\begin{figure}[htbp]
	\centering
    \includegraphics[scale=0.9]{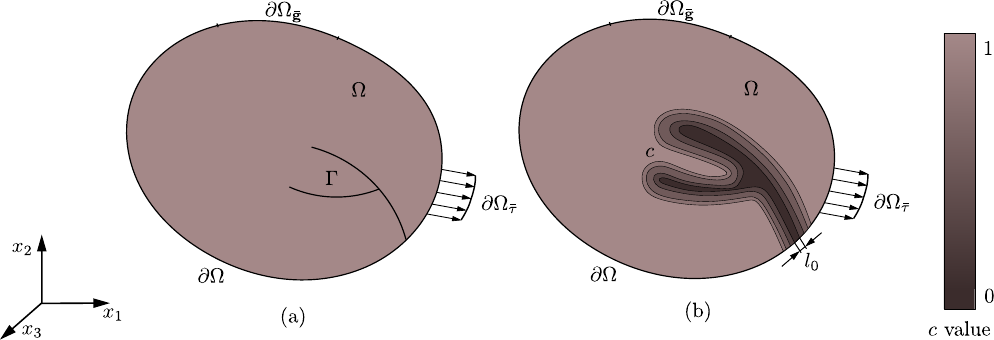}
	\caption[]{(a) Solid body $\Omega$ with crack path $\Gamma$ and (b) phase field approximation of the crack path $\Gamma$.}
    \label{fig:deform_bdy}
\end{figure}

An arbitrary deformable body occupying a domain $\Omega \subset \mathbb{R}^d$, where $d \in \{1,2,3\}$ denotes the spatial dimension, is considered, with outer boundary $\partial \Omega$, as illustrated in \FIG{fig:deform_bdy}. Appropriate kinematic constraints are imposed along the boundary of the domain $\partial \Omega_{\bar{g}} \subseteq \partial \Omega$. Although the formulation is presented in general form, only two-dimensional problems are considered in this work.

In accordance with Griffith’s theory of brittle fracture,\cite{Griffith1921} the total potential energy of the system under quasi-static conditions is defined as
\begin{equation}
\label{eqn:pot_eng}
\begin{aligned}
\Pi(\mathbf{u}, \Gamma) = \mathcal{W}^{\mathrm{int}}(\mathbf{u}, \Gamma) 
- \mathcal{W}^{\mathrm{ext}}(\mathbf{u}),
\end{aligned}
\end{equation}
where $\mathbf{u}(\mathbf{x})$ denotes the displacement field at material point $\mathbf{x}$. The term $\mathcal{W}^{\mathrm{int}}(\mathbf{u}, \Gamma)$ represents the internal energy stored in the body, which depends on the deformation and the presence of the crack $\Gamma$, while $\mathcal{W}^{\mathrm{ext}}(\mathbf{u})$ denotes the potential of the external mechanical work.

The external work potential is given by
\begin{equation}
\label{eqn:work_ext}
\begin{aligned}
\mathcal{W}^{\mathrm{ext}}(\mathbf{u}) 
= \int_{\partial \Omega_{\mathbf{\bar{\tau}}}} (\mathbf{\bar{\tau}} \cdot \mathbf{u})\, \mathrm{d}\partial \Omega_{\mathbf{\bar{\tau}}} 
+ \int_{\Omega} (\mathbf{f} \cdot \mathbf{u})\, \mathrm{d}\Omega,
\end{aligned}
\end{equation}
where $\bar{\mathbf{\tau}}$ denotes prescribed surface tractions acting on $\partial \Omega_{\bar{\mathbf{\tau}}}$, and $\mathbf{f}$ denotes the body force density.

The internal energy consists of the elastic strain energy $\Psi^{\mathrm{el}}$ and the fracture surface energy $\Psi^{\mathrm{f}}$,
\begin{equation}
\label{eqn:work_int}
\begin{aligned}
\mathcal{W}^{\mathrm{int}}(\mathbf{u}, \Gamma) 
= \Psi^{\mathrm{el}} + \Psi^{\mathrm{f}}.
\end{aligned}
\end{equation}

Within the phase-field framework,\cite{Bourdin2008} the fracture surface energy for anisotropic brittle fracture is approximated as \cite{Kakouris2018,Kakouris2019}
\begin{equation}
\label{eqn:frac_eng}
\begin{aligned}
\Psi^{\mathrm{f}} = \int_{\Gamma} \mathcal{G}_{c}(\theta) \, \mathrm{d}\Gamma 
\approx \int_{\Omega} \bar{\mathcal{G}}_{c} \, \mathcal{Z}_{c,\mathrm{Anis}} \, \mathrm{d}\Omega,
\end{aligned}
\end{equation}
where $\mathcal{G}_{c}(\theta)$ is the orientation-dependent critical energy release rate, $\bar{\mathcal{G}}_{c}$ is an effective fracture toughness, and $\mathcal{Z}_{c,\mathrm{Anis}}$ denotes the anisotropic crack density functional. The phase-field variable $c(\mathbf{x}) \in (0,1)$ describes the local material integrity, with $c=1$ corresponding to intact material and $c=0$ to fully fractured material.

The anisotropic crack density functional is defined as
\begin{equation}
\label{eqn:crack_dens_func}
\begin{aligned}
\mathcal{Z}_{c,\mathrm{Anis}} = \frac{(c-1)^{2}}{4 l_{0}} 
+ l_{0} \lvert \mathbf{\nabla} c \rvert^{2} 
+ l_{0}^{3} \sum_{ijkl} \gamma_{ijkl}  
\frac{\partial^{2} c}{\partial x_{i} \partial x_{j}} 
\frac{\partial^{2} c}{\partial x_{k} \partial x_{l}},
\end{aligned}
\end{equation}
where $l_{0} \in \mathbb{R}^{+}$ is the internal length-scale parameter and $\gamma_{ijkl}$ are the components of a dimensionless fourth-order tensor characterising anisotropic fracture behaviour.

For two-dimensional problems, $\gamma_{ijkl}$ is conveniently expressed in Voigt notation as
\begin{equation}
\label{eqn:gamma_tnr}
\begin{aligned}
\mathbf{\gamma} =
\begin{bmatrix}
\gamma_{1111} & \gamma_{1122} & \gamma_{1112} \\
\gamma_{2211} & \gamma_{2222} & \gamma_{2212} \\
\gamma_{1211} & \gamma_{1222} & \gamma_{1212}
\end{bmatrix}.
\end{aligned}
\end{equation}
In this representation, the components $\gamma_{ijkl}$ $(i,j,k,l = 1,2)$ correspond to the entries of the original fourth-order tensor written in matrix form. The tensor is normalised and dimensionless, ensuring that it only influences the directional dependence of the crack density functional without affecting its physical units. Its role is to weight the higher-order gradient contributions in \EQ{eqn:crack_dens_func}, thereby capturing the material’s anisotropic fracture response. In the isotropic limit, $\gamma_{ijkl}=0$, and the fracture process becomes direction-independent.

The elastic strain energy is given by
\begin{equation}
\label{eqn:elast_eng}
\begin{aligned}
\Psi^{\mathrm{el}} = \int_{\Omega} g(c) \psi^{\mathrm{el}}(\mathbf{\varepsilon}) \, \mathrm{d}\Omega,
\end{aligned}
\end{equation}
where $\psi^{\mathrm{el}}(\mathbf{\varepsilon})$ denotes the elastic strain energy density and $\mathbf{\varepsilon}$ is the symmetric strain tensor.  
Under the small-strain assumption, the strain tensor is given by $\mathbf{\varepsilon} = \tfrac{1}{2}\left( \nabla \mathbf{u} + \nabla \mathbf{u}^{T} \right)$, with $\nabla$ denoting the gradient operator.\cite{Bonet2008}

The function $g(c)$ in \EQ{eqn:elast_eng} is a \textit{degradation function} that governs the progressive loss of stiffness in the material as damage evolves. A commonly adopted quadratic degradation law is given by\cite{Miehe2010}
\begin{equation}
\label{eqn:degrad_func}
\begin{aligned}
g(c) = (1 - k_{\mathrm{f}})c^{2} + k_{\mathrm{f}},
\end{aligned}
\end{equation}
where $k_{\mathrm{f}}$ is a small positive model parameter introduced to prevent the complete loss of stiffness in the fully fractured regions and to ensure numerical stability. However, past works have shown that the parameter is not necessary for $\Gamma$-convergence,\cite{Borden2012} thus it is set to zero in this work.

\subsection{Influence of the anisotropic tensor} 
\label{sec:gamma_tnr}

The fourth-order tensor $\mathbf{\gamma}$ governs the anisotropic contribution to the fracture surface energy by weighting the higher-order gradient terms in \EQ{eqn:crack_dens_func}. Its components determine the directional dependence of the crack density functional and thus control the evolution of the phase field with respect to the crack orientation $\theta$. Different configurations of $\mathbf{\gamma}$ correspond to different material symmetry classes, including isotropic, cubic, and orthotropic behaviour.\cite{Ting1996}

The effect of crack orientation is introduced through a coordinate transformation. By rotating the global Cartesian coordinate system $\mathbf{x} = [x_{1}\; x_{2}]^{T}$ to a local system $\mathbf{x}_{\theta} = [x_{1\theta}\; x_{2\theta}]^{T}$, where $x_{1\theta}$ is tangential to the crack path $\Gamma$ and $x_{2\theta}$ is normal to it, the transformation
\begin{equation}
\label{eqn:rot_transf}
\begin{aligned}
\mathbf{x}_{\theta} = \mathbf{R}_{\theta} \, \mathbf{x}
\end{aligned}
\end{equation}
holds, with $\mathbf{R}_{\theta}$ denoting the standard two-dimensional rotation matrix (see Ref.~\citenum{Kakouris2019PhD}). This transformation enables the tensor $\mathbf{\gamma}$ to be rotated according to the crack orientation, thereby allowing evaluation of the directional dependence of the fracture surface energy density $\mathcal{G}_{c}(\theta)$.

\begin{figure}[htbp]
	\centering
    \begin{tabular}{c@{\hspace{4em}}c}
		\subfloat[\label{fig:gc_theta:a}]{
        \includegraphics[width=0.35\textwidth]{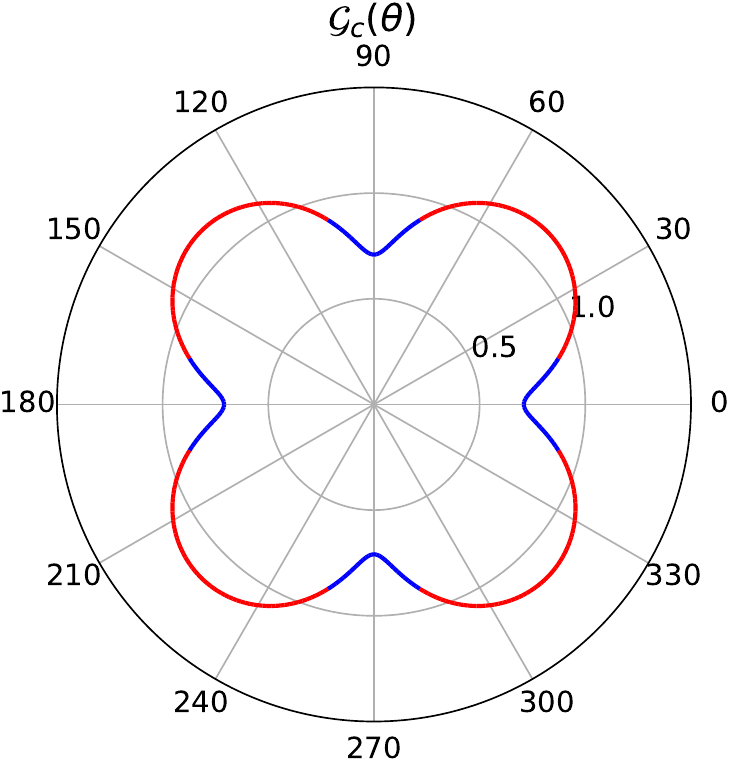}} &
		\subfloat[\label{fig:gc_theta:b}]{
			\includegraphics[width=0.35\textwidth]{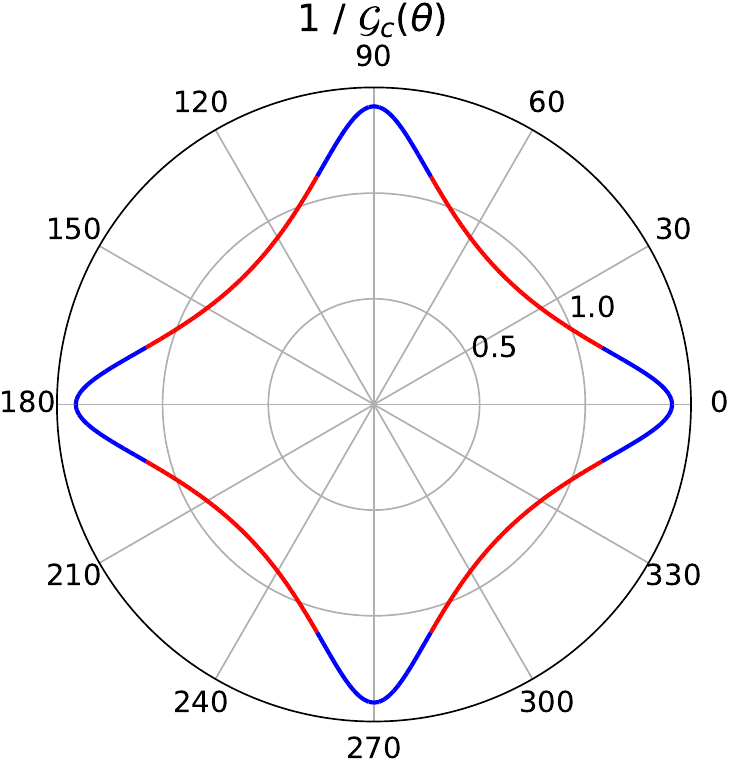}}
	\end{tabular}
\caption[]{\subref{fig:gc_theta:a} Polar representation of the fracture surface energy density $\mathcal{G}_{c}(\theta)$ and \subref{fig:gc_theta:b} its reciprocal $1/\mathcal{G}_{c}(\theta)$ for cubic material symmetry as a function of crack orientation $\theta$. The red portions of the polar plots denote the forbidden orientations associated with $\mathcal{S}^{r}(\theta) < 0$, while the blue portions indicate energetically stable paths with $\mathcal{S}^{r}(\theta) > 0$. $\mathcal{S}^{r}(\theta)$ is defined in Section \ref{sec:convex_nonconvex}.}
    \label{fig:gc_theta}
\end{figure}

For a given orientation $\theta$, the fracture surface energy density may be numerically evaluated as
\begin{equation}
\label{eqn:Gc_theta}
\begin{aligned}
\mathcal{G}_{c}(\theta) 
= \int_{\Gamma} \mathcal{G}_{c}(\theta) \, \mathrm{d}\Gamma
\approx \int_{-x_{lb}}^{+x_{lb}} 
\bar{\mathcal{G}}_{c} \, 
\mathcal{Z}_{c,\mathrm{Anis}} \, 
\mathrm{d}x_{2\theta},
\end{aligned}
\end{equation}
where $x_{lb}$ denotes a sufficiently large distance from the crack interface along the normal direction. A value of $x_{lb}=50 l_0$ is found to yield a reasonable approximation. This integration yields the orientation-dependent fracture toughness, which may be conveniently visualised in polar coordinates, as shown in \FIG{fig:gc_theta} for a cubic material symmetry. Such polar representations provide direct insight into the anisotropy induced by the tensor $\mathbf{\gamma}$.

For isotropic symmetry, all anisotropic parameters vanish ($\gamma_{ijkl}=0$), and the fracture surface energy density is identical in every direction, leading to a circular polar distribution of $\mathcal{G}_{c}(\theta)$. Cubic symmetry in two dimensions arises when $\gamma_{1111}=\gamma_{2222}\neq0$, $\gamma_{1122}=0$, and $\gamma_{1212}\neq0$ along the principal material axes, producing a fourfold polar pattern that reflects equal fracture resistance along orthogonal directions. 

Orthotropic symmetry represents a generalisation of cubic symmetry and is characterised in two dimensions by $\gamma_{1111}\neq\gamma_{2222}\neq0$. This form of anisotropy yields a twofold polar pattern, corresponding to distinct fracture resistance along the principal material directions. These variations in $\mathcal{G}_{c}(\theta)$ capture preferred and, in some cases, forbidden crack propagation directions that are characteristic of anisotropic materials. Figure~\ref{fig:gc_theta} illustrates the influence of $\mathbf{\gamma}$ on $\mathcal{G}_{c}(\theta)$ and its reciprocal for the symmetry classes considered.

\subsection{Convex and Non-convex reciprocals of surface energy density} 
\label{sec:convex_nonconvex}

Depending on the selected anisotropic parameters $\gamma_{ijkl}$, the surface energy density $\mathcal{G}_{c}(\theta)$ may exhibit either convex or non-convex behaviour. This distinction is of fundamental importance, as it determines whether all crack propagation directions are admissible or whether certain orientations are energetically forbidden. The convexity of $\mathcal{G}_{c}(\theta)$ may be examined by representing the surface energy density in polar coordinates (see \FIG{fig:gc_theta:a}).

A convex surface energy density corresponds to a \textit{weakly anisotropic} material, in which crack propagation is admissible in all directions. Conversely, a non-convex surface energy density indicates a \textit{strongly anisotropic} material, in which specific orientations become energetically unfavourable, leading to forbidden crack directions. These forbidden directions arise from a local loss of convexity of the surface energy.

The convexity condition can be quantified using the concept of \textit{surface stiffness},\cite{Li2015,Kakouris2019PhD} defined as
\begin{equation}
\label{eqn:surf_stiff}
\begin{aligned}
\mathcal{S}^{r}(\theta)
= \frac{\partial^{2} \mathcal{G}_{c}(\theta)}{\partial \theta^{2}}
+ \mathcal{G}_{c}(\theta),
\end{aligned}
\end{equation}
where $\mathcal{S}^{r}(\theta)$ characterises the stability of the fracture surface with respect to small perturbations in crack orientation. If $\mathcal{S}^{r}(\theta) > 0$, the surface energy is locally convex and the corresponding orientation is stable. In contrast, $\mathcal{S}^{r}(\theta) < 0$ indicates a loss of convexity and the presence of unstable, forbidden crack orientations.

From a physical standpoint, variations in surface energy during crack propagation are governed by $\mathcal{S}^{r}(\theta)$. When $\mathcal{S}^{r}(\theta) > 0$, the energy landscape remains locally stable and crack growth may proceed along that orientation. When $\mathcal{S}^{r}(\theta) < 0$, the energy landscape becomes locally unstable, and the crack path tends to deviate or bifurcate away from these orientations. These features are illustrated in \FIG{fig:gc_theta}, where red regions denote forbidden orientations and blue regions denote stable crack propagation paths.
\subsection{Strong form} \label{sec:strong_form}

The strong form of the coupled displacement-phase field problem is obtained from the variational formulation introduced in \EQ{eqn:pot_eng}. By invoking the principle of stationary potential energy, the first variation of the total energy functional with respect to the displacement field $\mathbf{u}$ and the phase field variable $c$ leads to the following governing balance equations;

\begin{equation}
\label{eqn:strong_form}
\left\{
\begin{aligned}
& \nabla \cdot \mathbf{\sigma} + \mathbf{f} = \mathbf{0}
&& \text{on } \Omega\text{,} && \text{Equilibrium} \\
& \prescript{(t-1)}{  }{ c }^{  }_{  } - c \ge 0 && \text{on } \Omega\text{,} && \text{Irreversibility} \\
& \left( \frac{4 l_{0} (1-k_{\mathrm{f}}) \psi^{\mathrm{el}}}{\bar{\mathcal{G}}_{c}} + 1 \right)c - 4 l_{0}^{2} \Delta c  + 4 l_{0}^{4} \sum_{ijkl} \gamma_{ijkl} \frac{\partial^{4}c}{\partial x_{i}\,\partial x_{j}\,\partial x_{k}\,\partial x_{l}} - 1 \ge 0 && \text{on } \Omega\text{,} && \text{Damage criterion} \\
& \left(\prescript{(t-1)}{  }{ c }^{  }_{  } - c\right)\left[\left( \frac{4 l_{0} (1-k_{f}) \psi^{\mathrm{el}}}{\bar{\mathcal{G}}_{c}} + 1 \right)c - 4 l_{0}^{2} \Delta c  + 4 l_{0}^{4} \sum_{ijkl} \gamma_{ijkl} \frac{\partial^{4}c}{\partial x_{i}\,\partial x_{j}\,\partial x_{k}\,\partial x_{l}} - 1 \right] = 0 && \text{on } \Omega\text{,} && \text{Loading-unloading}
\end{aligned}
\right.
\end{equation}
where $\mathbf{\sigma}$ denotes the Cauchy stress tensor and $\Delta(\cdot)$ denotes the Laplace operator.

In \EQ{eqn:strong_form}, the second inequality enforces crack irreversibility and prevents healing of the phase field. The third inequality defines the damage criterion governing the initiation and evolution of fracture, driven by the elastic energy density. The final equation corresponds to the Kuhn-Tucker loading-unloading conditions and ensures consistency between the irreversibility constraint and the damage criterion. Further details on how irreversibility is applied in this work are provided in Section \ref{sec:irreversibility}. All quantities without the superscript $(t\!-\!1)$ are evaluated at the current pseudo-time step $t$.

\EQ{eqn:strong_form} is supplemented by the following initial and boundary conditions.
\begin{equation}
\label{eqn:strong_form_cond}
\left\{
\begin{aligned}
& \mathbf{u} = \prescript{(0)}{  }{ \mathbf{u} }^{  }_{  }, && \text{on } \prescript{(0)}{  }{ \Omega }^{  }_{ } \\
& \mathbf{u} = \mathbf{\bar{g}}, && \text{on } \partial \Omega_{\mathbf{\bar{g}}} \\
& \mathbf{\sigma}\mathbf{n} = \mathbf{\bar{\tau}}, && \text{on } \partial \Omega_{\mathbf{\bar{\tau}}} \\
& c = \prescript{(0)}{  }{ c }^{  }_{  }, && \text{on } \prescript{(0)}{  }{ \Omega }^{  }_{ } \\
& \Bigg[ 4 l_{0}^{2} \nabla c - 2 l_{0}^{4} \sum_{ijkl} \gamma_{ijkl}  \left( \frac{\partial^{3}c}{\partial x_{j}\partial x_{k}\partial x_{l}} \right)  - 2 l_{0}^{4} \sum_{ijkl} \gamma_{ijkl} \left( \frac{\partial^{3}c}{\partial x_{i}\partial x_{j}\partial x_{k}} \right) \Bigg] \cdot \mathbf{n} = 0, && \text{on } \partial\Omega \\
& 2 l_{0}^{4} \sum_{ijkl} \gamma_{ijkl} \left( \frac{\partial^{2}c}{\partial x_{k}\partial x_{l}} \right) + 2 l_{0}^{4} \sum_{ijkl} \gamma_{ijkl} \left( \frac{\partial^{2}c}{\partial x_{i}\partial x_{j}} \right) = 0. && \text{on } \partial\Omega
\end{aligned}
\right.
\end{equation}
where $\mathbf{n}$ is the unit outward normal vector on the boundary.

\subsection{Elastic strain energy density} \label{sec:elc_strn_eng_dens}

The constitutive response of the solid is described by a general linear elastic anisotropic material model. In two dimensions, the fourth-order elastic stiffness tensor is expressed in Voigt notation as
\begin{equation}
\label{eqn:elc_const_mtrx}
\begin{aligned}
\mathbf{D}^{\mathrm{el}} =
\begin{bmatrix}
D^{\mathrm{el}}_{1111} & D^{\mathrm{el}}_{1122} & D^{\mathrm{el}}_{1112} \\
D^{\mathrm{el}}_{2211} & D^{\mathrm{el}}_{2222} & D^{\mathrm{el}}_{2212} \\
D^{\mathrm{el}}_{1211} & D^{\mathrm{el}}_{1222} & D^{\mathrm{el}}_{1212}
\end{bmatrix}.
\end{aligned}
\end{equation}

Material degradation due to fracture is introduced through the scalar degradation function $g(c)$, such that the effective stiffness tensor reads
\begin{equation}
\label{eqn:const_mtrx}
\begin{aligned}
\mathbf{D} = g(c)\, \mathbf{D}^{\mathrm{el}}.
\end{aligned}
\end{equation}
The Cauchy stress tensor is given by
\begin{equation}
\label{eqn:cauchy_tnr}
\begin{aligned}
\mathbf{\sigma} = \mathbf{D} : \mathbf{\varepsilon}.
\end{aligned}
\end{equation}
The elastic strain energy density is defined as
\begin{equation}
\label{eqn:elc_strn_eng_dens}
\begin{aligned}
\psi^{\mathrm{el}}(\mathbf{\varepsilon}) = \frac{1}{2}\, \mathbf{\varepsilon} : \mathbf{D}^{\mathrm{el}} : \mathbf{\varepsilon}.
\end{aligned}
\end{equation}

In contrast to many phase-field formulations that employ a tensile-compressive or spectral decomposition of the elastic energy \cite{Ambati2015}, no energetic split is adopted in the present work. Instead, a single, unsplit elastic strain energy density is used for general anisotropic elasticity. This modelling choice is sufficient for the benchmark problems considered herein and does not affect the qualitative or quantitative conclusions of this study. Accordingly, further examination of alternative energy splits is not pursued, but will be considered in future work.

\subsection{Irreversibility} \label{sec:irreversibility}

Irreversibility is expressed by the second inequality in \EQ{eqn:strong_form}. It enforces the physical requirement that material degradation is monotonic, such that once damage has occurred, the material cannot recover its original integrity and crack healing is precluded. Several strategies exist to enforce irreversibility, including history-field formulations,\cite{Miehe2010} variational inequality formulations based on constrained energy minimisation,\cite{Bourdin2008} penalty-based regularisation,\cite{Gerasimov2019} and augmented Lagrangian techniques.\cite{Wheeler2014}

Among the available strategies, the history-field formulation is widely used. In this approach, the crack driving force is taken as the maximum value attained by the elastic strain energy density over the loading history. This formulation has been successfully employed in both finite element studies \cite{Miehe2010} and NN-based approaches.\cite{Goswami2020a,Goswami2020b,Motlagh2023} In contrast, Manav \emph{et al.}\cite{Manav2024} adopt a penalty-based approach in order to retain a fully variational formulation compatible with the DRM.

A penalty-based approach is also adopted in the present work. Irreversibility is \textit{weakly} enforced by augmenting the total potential energy, \EQ{eqn:pot_eng}, with an additional penalty term of the form
\begin{equation}
\label{eqn:irr_eng}
\begin{aligned}
\mathcal{P}^{\mathrm{irr}} = \int_{\Omega} \frac{1}{2} \kappa_{\mathrm{irr}} \langle \prescript{(t-1)}{  }{ c }^{  }_{  } - c \rangle_-^2\, \mathrm{d}\Omega,
\end{aligned}
\end{equation}
such that
\begin{equation}
\label{eqn:pot_eng_irr}
\begin{aligned}
\Pi(\mathbf{u}, c) = \mathcal{W}^{\mathrm{int}}(\mathbf{u}, c) 
- \mathcal{W}^{\mathrm{ext}}(\mathbf{u}) + \mathcal{P}^{\mathrm{irr}} (c).
\end{aligned}
\end{equation}
The operator $\langle x \rangle_- = \min(x,0)$ denotes the negative-part (Macaulay) operator and ensures that only violations of the irreversibility constraint are penalised.

Following the work by Gerasimov and De Lorenzis,\cite{Gerasimov2019} the penalty parameter $\kappa_{\mathrm{irr}}$ is selected using AT-2 scaling. Noting that the isotropic part of the adopted crack density functional $\mathcal{Z}_{c,\mathrm{Anis}}$ corresponds to an AT-2 formulation with an effective length scale $2l_0$, the penalty parameter is chosen as
\begin{equation}
\label{eqn:irr_coeff}
\begin{aligned}
\kappa_{\mathrm{irr}} = \frac{\bar{\mathcal{G}}_{c}}{2 l_{0}}\left(\frac{1}{\mathrm{tol}_{\mathrm{irr}}^{2}} - 1\right).
\end{aligned}
\end{equation}

The irreversibility tolerance satisfies $0 < \mathrm{tol}_{\mathrm{irr}} \le 1$; in this work, $\mathrm{tol}_{\mathrm{irr}} = 2 \times 10^{-2}$ is adopted.

\subsection{Non-dimensionalisation scheme} \label{sec:non_dims}

A non-dimensionalisation scheme is employed to improve numerical conditioning and to ensure balanced scaling of the governing equations prior to their solution using the DRM. The present non-dimensionalisation strategy follows the approach of Manav \emph{et al.}\cite{Manav2024} with appropriate modifications to account for anisotropic elasticity.

Let $L$ denote a characteristic length of the body. Spatial coordinates and the phase-field length scale are rescaled as
\begin{equation}
\begin{aligned}
\tilde{\mathbf{x}} = \frac{\mathbf{x}}{L}, 
\qquad
\tilde{l}_0 = \frac{l_0}{L}.
\end{aligned}
\end{equation}
A reference elastic modulus $E_{\mathrm{ref}}$ is introduced to normalise the anisotropic stiffness tensor. In this work, $E_{\mathrm{ref}}$ is chosen as a representative magnitude of the elastic stiffness, e.g. the maximum diagonal component of $\mathbf{D}^{\mathrm{el}}$. The displacement field is scaled according to
\begin{equation}
\begin{aligned}
\tilde{\mathbf{u}}
=
\frac{\mathbf{u}}{L}
\left(
\frac{\bar{\mathcal{G}}_{c}}{E_{\mathrm{ref}}\,\ell_{\mathrm{eff}}}
\right)^{-1/2},
\end{aligned}
\end{equation}
where $\ell_{\mathrm{eff}} = 2l_0$ denotes the effective length scale associated with the isotropic part of the adopted crack density functional.

The symmetric strain tensor is correspondingly scaled as
\begin{equation}
\begin{aligned}
\tilde{\mathbf{\varepsilon}}
=
\mathbf{\varepsilon}
\left(
\frac{\bar{\mathcal{G}}_{c}}{E_{\mathrm{ref}}\,\ell_{\mathrm{eff}}}
\right)^{-1/2}.
\end{aligned}
\end{equation}
The dimensionless anisotropic stiffness tensor is defined as
\begin{equation}
\begin{aligned}
\tilde{\mathbf{D}}^{el} = \frac{\mathbf{D}^{el}}{E_{\mathrm{ref}}},
\end{aligned}
\end{equation}
such that the non-dimensional elastic strain energy density becomes
\begin{equation}
\begin{aligned}
\tilde{\psi}^{\mathrm{el}}
=
\frac{\ell_{\mathrm{eff}}}{\bar{\mathcal{G}}_{c}}\,\psi^{\mathrm{el}}
=
\frac{1}{2}\,
\tilde{\mathbf{\varepsilon}}
:
\tilde{\mathbf{D}}^{\mathrm{el}}
:
\tilde{\mathbf{\varepsilon}}.
\end{aligned}
\end{equation}
The phase-field variable $c$ is already dimensionless and is therefore unchanged. Gradients are rescaled as
\begin{equation}
\begin{aligned}
\tilde{\nabla} = L \nabla,
\qquad
\partial_{ij} = L^{-2}\tilde{\partial}_{ij}.
\end{aligned}
\end{equation}
With these definitions, the anisotropic crack density functional $\mathcal{Z}_{c,\mathrm{Anis}}$ and the total potential energy can be expressed in fully non-dimensional form.

Notably, in contrast to isotropic elasticity, the non-dimensional energy in the present formulation depends on the dimensionless anisotropic stiffness tensor $\tilde{\mathbf{D}}^{\mathrm{el}}$ and on the dimensionless length scale $\tilde{l}_0$, in addition to the anisotropic gradient coefficients $\gamma_{ijkl}$. In the isotropic limit, $\tilde{\mathbf{D}}^{\mathrm{el}}$ depends only on Poisson’s ratio. For notational convenience, all quantities are henceforth assumed to be non-dimensional, and the tilde notation is omitted.

\section{Extending the Deep Ritz method for anisotropic fracture} \label{sec:ext_deep_ritz_method}

\subsection{Deep Ritz method} \label{sec:deep_ritz_method}

The DRM is a variational, physics-informed learning framework for solving boundary value problems by minimising an energy functional using deep NNs.\cite{Yu2018} In contrast to residual-based approaches, such as PINNs,\cite{Raissi2019} the DRM operates directly at the level of the variational formulation and seeks approximations of the solution fields that minimise the total potential energy. This characteristic makes the DRM particularly suitable for phase-field fracture problems, where the governing equations follow from stationarity of a non-convex energy functional. The effectiveness of the DRM for second-order isotropic phase-field models has been demonstrated by \cite{Manav2024} However, extension to higher-order anisotropic phase-field fracture formulations is not straightforward.  The anisotropic crack density functional introduced in Section \ref{sec:phase_field_anisotr} involves additional gradient terms and direction-dependent surface energies, which significantly increase the non-convexity of the variational problem and impose stricter regularity requirements on the approximation space. 
These features fundamentally alter both the mathematical structure of the governing functional and the associated optimisation landscape.

Within this framework, the solution fields are represented by feed-forward NNs, consisting of successive affine transformations and scalar non-linear activation functions. The displacement field $\mathbf{u}$ and the phase-field variable $c$ are approximated as
\begin{equation}
\label{eqn:nn_ansatz}
\begin{aligned}
\mathbf{u}(\mathbf{x}) \approx \mathbf{u}_{\eta}(\mathbf{x}),
\qquad
c(\mathbf{x}) \approx c_{\eta}(\mathbf{x}),
\end{aligned}
\end{equation}
where $\eta$ denotes the set of trainable network parameters.

\begin{figure}[htbp]
	\centering
    \includegraphics[width=\textwidth]{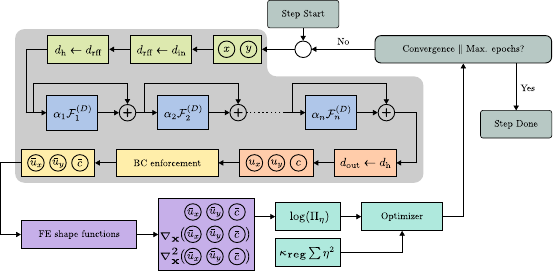}
    \caption[]{Schematic illustration of the DRM employed in this work. Spatial coordinates are input to the NN to obtain approximations of the displacement- and phase-fields. These are used within a quadrature scheme to evaluate the total potential energy, which is subsequently minimised by an optimiser until a convergence criterion is reached. Parameters of the network are explained in Section \ref{sec:deep_neural_network}.}
    \label{fig:drm_schematic}
\end{figure}

The coupled phase-field fracture problem is then recast as the following optimisation problem:
\begin{equation}
\label{eqn:drm_opt}
\begin{aligned}
\min_{\eta}
\; \Pi\big(\mathbf{u}_{\eta},\,c_{\eta}\big),
\end{aligned}
\end{equation}
where $\Pi$ is the non-dimensional total potential energy defined in \EQ{eqn:pot_eng_irr}, including elastic, fracture, and irreversibility contributions. A schematic illustration of the overall DRM workflow adopted in this work is shown in \FIG{fig:drm_schematic}.
In practice, a loss function is constructed from the energy functional, and the network parameters are determined by minimising this loss using gradient-based optimisation algorithms. The loss function utilised in this work is discussed in Section \ref{sec:loss_function}.

Numerical integration of the energy functional is performed over a B-spline-based spatial discretisation of the domain,\cite{deBoor2001,Hughes2005} following the quadrature strategy reported by \cite{Manav2024} The NN provides values of the displacement and phase-field variables at the B-spline control points. Field gradients and higher-order derivatives required by the energy functional are computed using the adopted B-spline basis functions and their derivatives at quadrature points.

This hybrid neural-spline strategy combines the expressive approximation capability of NNs with the numerical robustness of spline-based quadrature. 
In particular, it permits stable evaluation of higher-order spatial derivatives while avoiding potential instabilities associated with automatic differentiation in strongly non-convex settings. 
The governing equations are enforced implicitly through variational minimisation of the total potential energy.

After training, the NN delivers continuous approximations of the displacement and phase-field variables over the domain. These approximations are subsequently used to evaluate strains, stresses, and crack patterns.

\subsection{Loss function} \label{sec:loss_function}

The loss function employed for network training is defined as
\begin{equation}
\label{eqn:loss_log_energy}
\begin{aligned}
\mathcal{L}
=
\log\!\left(\Pi_{\eta}\right).
\end{aligned}
\end{equation}
The logarithmic transformation is introduced to improve numerical conditioning and to ensure that the magnitude of the loss remains of order unity throughout training. This facilitates the use of consistent stopping criteria and regularisation parameters across different problems and loading stages.

\subsection{Deep Neural Network} \label{sec:deep_neural_network}

This subsection presents the adopted NN architecture, activation functions, boundary condition enforcement strategy, and key implementation aspects.

\subsubsection{Network architecture} \label{sec:network_architecture}

The DRM formulation is naturally suited to monolithic energy minimisation. The NN therefore predicts the displacement and phase fields simultaneously. Although physically distinct fields may in principle be represented by independent NNs, it has been demonstrated by Manav \emph{et al.}\cite{Manav2024} that employing a single network improves the learning of crack nucleation due to enhanced coupling between the displacement and phase fields. Independent networks are therefore not explored in the present work. This monolithic representation may permit the use of larger displacement increments than those typically adopted in staggered finite-element schemes, where the fields are solved in an alternating manner.\cite{Bharali2022,Jin2024} The potential computational efficiency gain is accompanied by increased optimisation complexity.

The input to the network consists of the spatial coordinates of the B-spline control points $\mathbf{x}$, while the outputs correspond to $(\mathbf{u}, c)$. For the two-dimensional problems considered herein, the input dimension is two and the output dimension is three. For clarity, the notation used in the schematic of Figure~\ref{fig:drm_schematic} is briefly summarised here.  The symbol $d_\text{in}$ denotes the input dimension of the NN, corresponding to the spatial coordinates $\mathbf{x}$.  The RFF mapping increases the dimensionality of the input representation, producing a feature vector of dimension $d_\text{rff}=2m$.  This feature vector is subsequently projected to the hidden layer dimension $d_h$, which corresponds to the width of the residual network.  Finally, the output layer maps the hidden representation to the physical fields $(\mathbf{u},c)$ with output dimension $d_\text{out}$, which equals three for the two-dimensional problems considered in this work.

A ReZero architecture embedded within a residual network (ResNet) framework is adopted.\cite{He2015,Bachlechner2020} Residual connections facilitate gradient flow and alleviate vanishing gradient effects in deep networks, enabling stable training at increased depth. Such stability is essential in the present work, as the minimisation of a non-convex phase-field energy functional requires expressive neural representations. The ReZero modification introduces learnable residual scaling parameters initialised at zero. This ensures that each residual block initially behaves as an identity mapping. The network therefore starts from a stable configuration and progressively learns non-trivial transformations. This property is particularly advantageous in incremental loading schemes, where the network parameters from the previous load step provide a near-identity initialisation for the subsequent energy minimisation.

The network is organised into a sequence of residual blocks. Let $\mathcal{S}_0 \in \mathbb{R}^{n}$ denote the feature representation obtained from the input layer, where $n$ is the hidden layer dimension. The output of residual block $I$ is defined as
\begin{equation}
\label{eqn:out_res_block}
\mathcal{S}_{I} = \alpha_{I} \mathcal{F}_{I}^{(\mathcal{D})}(\mathcal{S}_{I-1}) + \mathcal{S}_{I-1},
\end{equation}
where $\mathcal{S}_{I-1}$ denotes the input to the block, $\mathcal{F}_{I}^{(\mathcal{D})}$ represents a feed-forward subnetwork of depth $\mathcal{D}$, and $\alpha_{I}$ is a learnable residual scaling coefficient. In the ReZero formulation, $\alpha_{I}$ is initialised at zero, such that each block initially acts as an identity mapping.
For any feature vector $\mathcal{X} \in \mathbb{R}^{n}$, the internal mapping $\mathcal{F}_{I}^{(\mathcal{D})}$ is constructed recursively as
\begin{equation}
    \begin{gathered}
        \mathcal{F}_{I}^{(\mathcal{D})}(\mathcal{X})=h_{I}^{(\mathcal{D})}, \\
        h_{I}^{(k)} = \phi\!\left(\mathbf{w}_{I,k} h_{I}^{(k-1)} + \mathbf{b}_{I,k}\right), \quad k=1,\ldots,\mathcal{D}, \\
        h_{I}^{(0)} = \mathcal{X}.
    \end{gathered}
\label{eqn:int_mapping}
\end{equation}
Here, $\mathbf{w}_{I,k}$ and $\mathbf{b}_{I,k}$ denote the weights and biases of layer $k \in \{1,\ldots,\mathcal{D}\}$ for block $I$, and $\phi$ denotes the activation function. The specific choice of activation function is discussed in Section \ref{sec:activation_function}.

The input coordinates are not supplied directly to the residual network. Instead, they are first mapped to a higher-dimensional feature space using 
RFF, which improves the representation of high-frequency  components in the displacement and phase fields. \cite{Tancik2020} Such enrichment is particularly beneficial in fracture problems, where the phase field localises over a small internal length scale and the displacement field exhibits high spatial gradients in the vicinity of crack tips. Standard multilayer perceptrons are known to display spectral bias towards low-frequency functions.\cite{Rahaman2019} The Fourier feature mapping mitigates this limitation by enabling the network to represent higher-wavenumber components more effectively.\cite{Tancik2020}

Given spatial coordinates $\mathbf{x} \in \mathbb{R}^{d}$, the RFF mapping 
$\mathcal{S}_{\mathrm{rff}} : \mathbb{R}^{d} \rightarrow \mathbb{R}^{2m}$ is defined as
\begin{equation}
\label{eqn:ref_mapping}
\begin{aligned}
\mathcal{S}_{\mathrm{rff}}(\mathbf{x}) =
\mathcal{S}_{\mathrm{0}}(\mathbf{x}) =
\left[
\cos(\mathbf{v}_1^\top \mathbf{x}), \ldots, \cos(\mathbf{v}_m^\top \mathbf{x}),
\sin(\mathbf{v}_1^\top \mathbf{x}), \ldots, \sin(\mathbf{v}_m^\top \mathbf{x})
\right]^\top,
\end{aligned}
\end{equation}
where $\mathbf{v}_{J} \in \mathbb{R}^{d}$, $J=1,\ldots,m$, are randomly sampled 
wave vectors. Each vector $\mathbf{v}_{J}$ determines a spatial frequency through 
its magnitude $\|\mathbf{v}_{J}\|$ and an orientation through its direction. The vectors are drawn independently from a multivariate normal distribution,
\begin{equation}
\label{eqn:norm_dist_vec}
\begin{aligned}
\mathbf{v}_{J} \sim \mathcal{N}(\mathbf{0}, \sigma_{\mathrm{rff}}^{2} \mathbf{I}),
\end{aligned}
\end{equation}
where $\mathcal{N}(\mathbf{0}, \sigma_{\mathrm{rff}}^{2} \mathbf{I})$ denotes a zero-mean Gaussian distribution with covariance matrix $\sigma_{\mathrm{rff}}^{2} \mathbf{I}$. The parameter $\sigma_{\mathrm{rff}}$ therefore represents the standard deviation of the sampled wave vectors and controls the effective bandwidth of the induced feature space. The parameters $\sigma_{\mathrm{rff}} = 0.1$ and $m = 192$ are adopted in all simulations of this work. 

The RFF features are subsequently projected to the hidden dimension $n$ 
through a linear transformation before entering the first residual block. 
A corresponding linear projection is applied at the output layer to map 
the final feature representation to the physical fields $(\mathbf{u}, c)$.

All linear layers are initialised using Glorot initialisation,\cite{Glorot2010} while the residual scaling coefficients $\alpha_I$ are initialised at zero  in accordance with the ReZero formulation. This ensures stable initial training dynamics and consistent behaviour within the incremental loading scheme.

\subsubsection{Activation function} \label{sec:activation_function}

There are multiple options for activation function. The rectified linear unit (ReLU) activation function was initially considered, motivated by theoretical and practical studies in the literature.\cite{MullerZeinhofer2020, Manav2024} ReLU is widely adopted due to its computational efficiency and favourable optimisation behaviour. However, satisfactory training was observed to require sufficiently large network widths. For narrower architectures, convergence deteriorated and, in certain cases, optimisation stagnated. This behaviour is consistent with the well-known ``dying ReLU'' phenomenon, whereby neurons may become inactive when the input remains in the negative half-space, resulting in vanishing gradients and limited recovery.\cite{Lu2020} In addition, ReLU generates piecewise linear outputs and therefore does not naturally represent smooth second-order variation in the approximated fields. In variational formulations involving higher-order gradient terms, such limited smoothness may restrict accurate modelling of curvature-dependent energy contributions.

The hyperbolic tangent activation function, $\tanh$, was therefore considered as an alternative. This activation is frequently employed in PINN formulations.\cite{Raissi2019, Motlagh2023, Goswami2020a} The $\tanh$ function is smooth and infinitely differentiable. It is symmetric about the origin and bounded such that $|\tanh(\mathbf{z})| \le 1$ for all $\mathbf{z}$. These properties promote stable optimisation and yield smooth field representations compatible with higher-order variational formulations.

\begin{figure}[htbp]
\centering
\begin{tabular}{cc}
\subfloat[\label{fig:activation_function_scaling}]{
\includegraphics[width=0.45\textwidth]{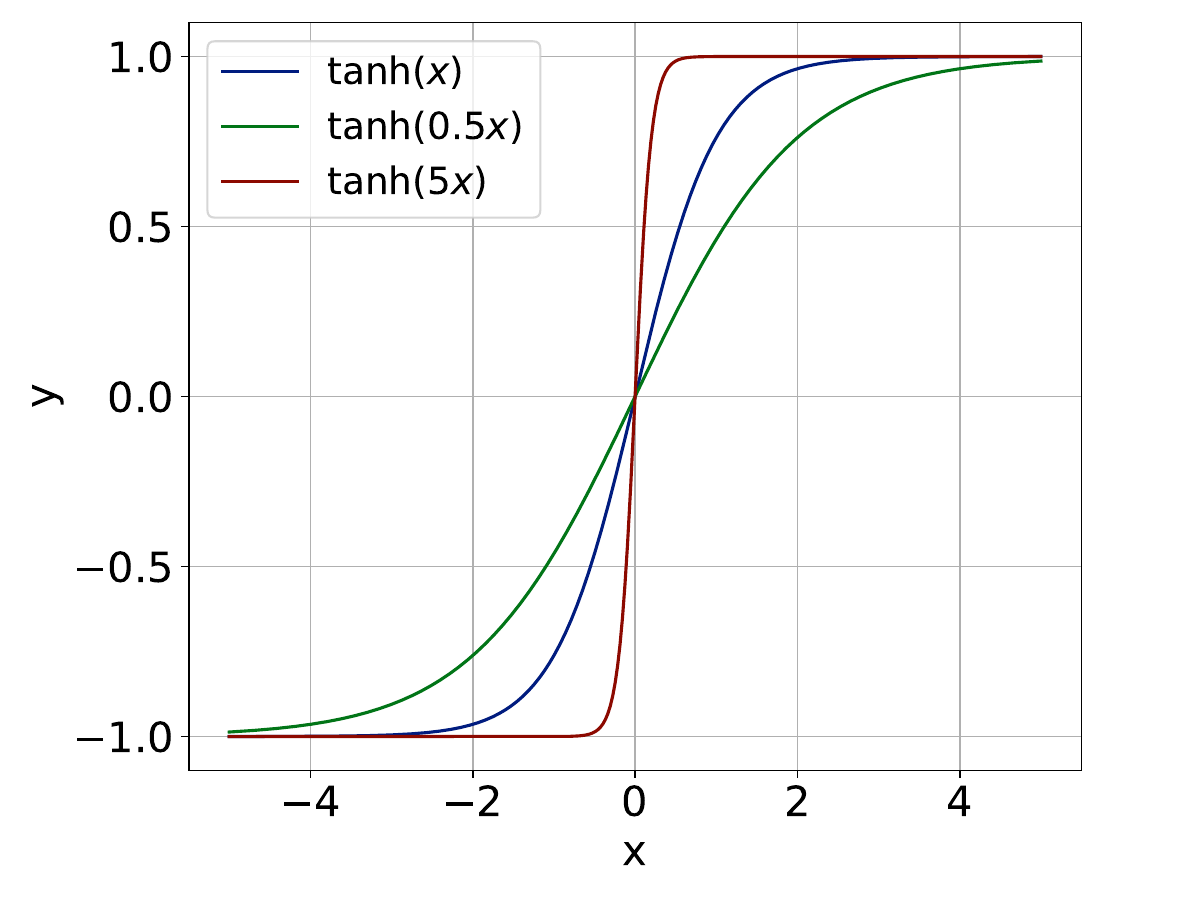}} &
\subfloat[\label{fig:nonsmooth_sigmoid}]{
\includegraphics[width=0.45\textwidth]{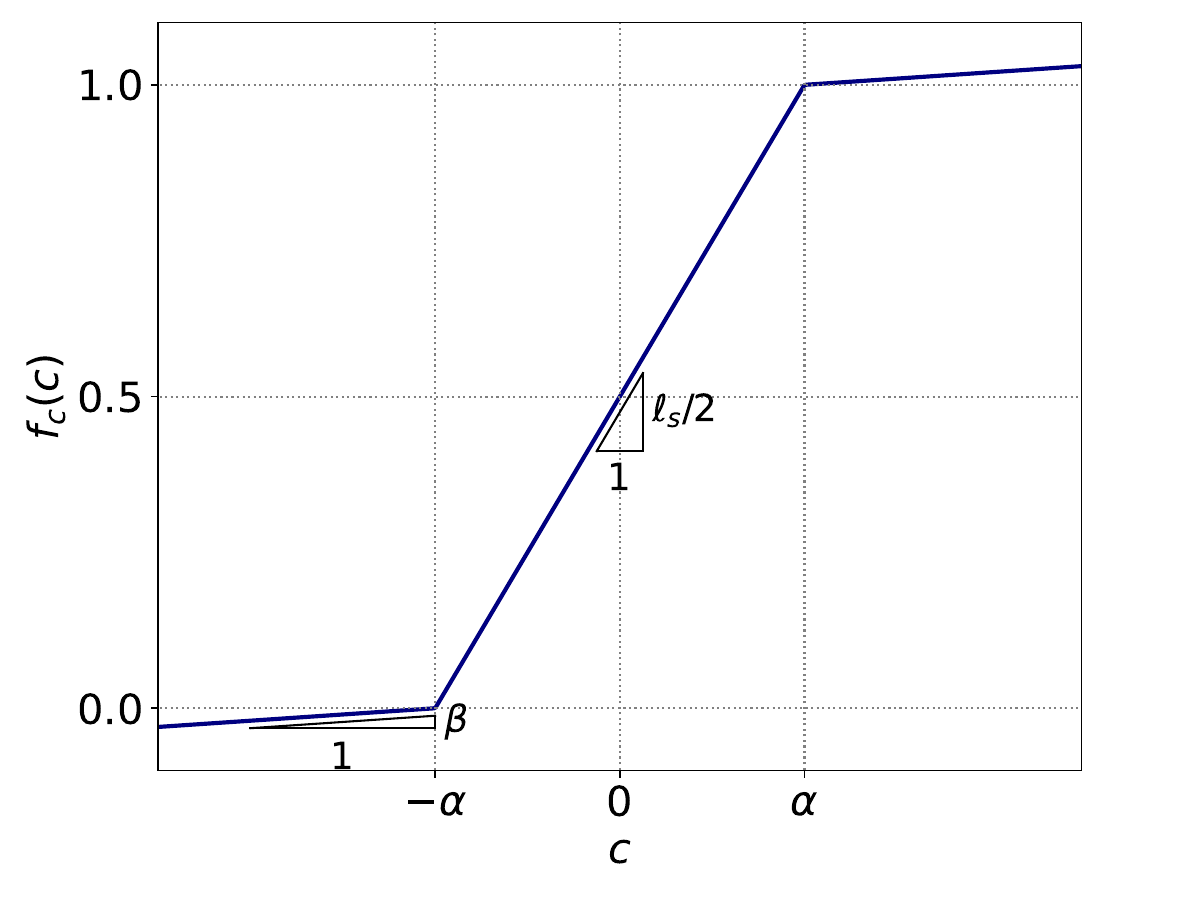}}
\end{tabular}
\caption[]{\subref{fig:activation_function_scaling} Effect of the scaling coefficient $r$ on the $\tanh$ activation function. Smaller values extend the quasi-linear regime. \subref{fig:nonsmooth_sigmoid} Non-smooth piecewise constraint used to enforce admissible phase-field values.}
\label{fig:activation_and_pf_constraint}
\end{figure}

A known limitation of $\tanh$ is saturation for large $|\mathbf{z}|$, where the derivative approaches zero. This may lead to vanishing gradient effects during optimisation. To mitigate this behaviour, a learnable scaling coefficient is introduced and the activation function is defined as
\begin{equation}
\label{eqn:tanh_func}
\begin{aligned}
\phi(\mathbf{z}) \coloneqq \phi(r \mathbf{z}) = \tanh(r \mathbf{z}).
\end{aligned}
\end{equation}
The parameter $r$ is introduced as a trainable scalar coefficient and forms part of the set of network parameters. It is initialised prior to optimisation and updated during training together with the weights and biases. The coefficient $r$ is shared across all layers and controls the effective slope of the activation function, thereby adapting the width of the near-linear regime during optimisation. Smaller values of $r$ extend the quasi-linear region, whereas larger values increase nonlinearity. A schematic representation of the scaled activation function and the effects of the scaling coefficient $r$ is presented in \FIG{fig:activation_function_scaling}.

\subsubsection{Weight regularisation} \label{sec:weight_regularisation}

Network regularisation is commonly employed to improve generalisation performance and reduce overfitting. In the context of the DRM applied to fracture mechanics, Manav \emph{et al.}\cite{Manav2024} demonstrated that weight regularisation plays a critical role in obtaining physically consistent solutions. In phase-field fracture models, the displacement field may exhibit sharp spatial variations in the vicinity of crack tips. Within a NN approximation, such steep gradients can lead to large strain magnitudes and unstable optimisation behaviour.

In classical finite element formulations, these effects may be alleviated through mesh refinement. In NN-based approximations, however, refinement of the discretisation is not directly applicable. Instead, regularisation of the trainable parameters provides a mechanism for controlling excessive parameter growth and stabilising optimisation.

Accordingly, the loss functional introduced in \EQ{eqn:loss_log_energy} is augmented by a L2 regularisation penalty on the network weights. The regularised loss is defined as
\begin{equation}
\label{eqn:reg_loss_func}
\begin{aligned}
\mathcal{L}_{\mathrm{reg}} = \log(\Pi_\eta) + \kappa_{\mathrm{reg}} \sum_{I,J} \|\mathbf{w}_{I,J}\|_F^2,
\end{aligned}
\end{equation}
where 
$\kappa_{\mathrm{reg}} = 10^{-5}$ and $\|\cdot\|_F$ denotes the Frobenius norm. The regularisation term acts solely on the network parameters and does not modify the underlying physical energy functional.

\subsubsection{Boundary condition enforcement}\label{sec:boundary_condition_enforcement}

In PINNs, boundary conditions are commonly imposed through additional penalty terms in the loss functional.\cite{Raissi2019} This approach corresponds to a weak imposition strategy, whereby boundary compliance is encouraged but not satisfied identically. For variational formulations, such penalty-based enforcement may lead to suboptimal convergence behaviour, particularly when the boundary terms interact strongly with the energy functional.\cite{Yu2018}

In variational fracture modelling using NNs, strong imposition of boundary conditions is frequently adopted through architectural modification.\cite{Motlagh2023,Goswami2020a,Manav2024} In this approach, admissibility is enforced by construction. The present work follows this strategy by incorporating distance-based shape functions to ensure exact satisfaction of prescribed boundary and loading conditions.

\paragraph{Displacement field}

Let $d_F(\mathbf{x})$ and $d_L(\mathbf{x})$ denote smooth distance functions associated with fixed and loaded boundary regions, respectively. These functions vanish outside their prescribed supports and are constructed to vanish on the complementary regions and remain strictly positive within the prescribed boundary supports.\cite{Sukumar2022} Let $\mathbf{u}(\mathbf{x}) \in \mathbb{R}^2$ denote the displacement field predicted by the NN. The augmented displacement field is defined as
\begin{equation}
\label{eqn:presc_disp}
\begin{aligned}
\bar{\mathbf{u}}(\mathbf{x})
=
\left[\left(1 - d_F(\mathbf{x})\right)\left(1 - d_L(\mathbf{x})\right)\mathbf{u}(\mathbf{x})
+
d_L(\mathbf{x})\right]\,\mathbf{U}_p,
\end{aligned}
\end{equation}
where $\mathbf{U}_p \in \mathbb{R}^2$ denotes the prescribed displacement vector. This construction enforces homogeneous displacement on the fixed boundary and the prescribed displacement on the loaded boundary exactly. The supports of $d_F(\mathbf{x})$ and $d_L(\mathbf{x})$ are assumed disjoint.

\begin{figure}[htbp]
\centering
\includegraphics[width=0.90\textwidth]{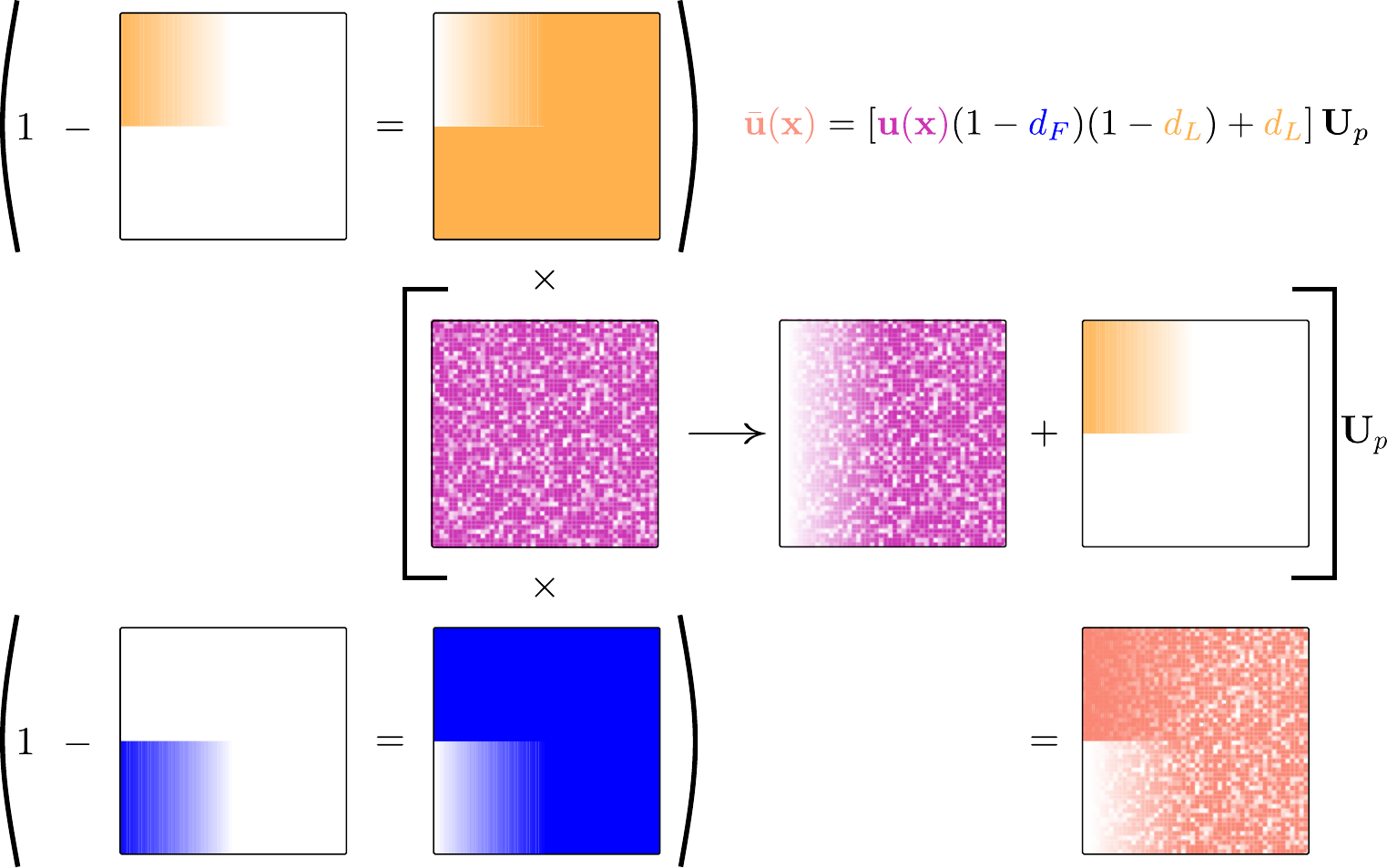}
\caption{Displacement field enforcement via distance functions. The shaded regions represent non-zero values, while the white regions are zero. The yellow field represents the loading distance function, indigo represents the fixed boundary conditions, and magenta represents the NN predictions, all scale $\sim1$. The light orange field represents the output scaled by the prescribed displacement, satisfying the prescribed loading and boundary conditions. The regions bounded by both the loaded and the fixed distance functions are enforced to be zero by inverting and composing with the network prediction. The exact prescribed displacement is then obtained by reintroducing the loaded area and scaling by the prescribed displacement $\mathbf{U}_p$.}
\label{fig:displacement_field_enforcement}
\end{figure}

For two-dimensional problems with loading angle $\omega$, the prescribed displacement vector is defined as
\begin{equation}
\label{eqn:disp_mag}
\begin{aligned}
\mathbf{U}_p = U_0 \begin{bmatrix} \cos(\omega) \\ \sin(\omega) \end{bmatrix},
\end{aligned}
\end{equation}
where $U_0$ is the prescribed displacement magnitude. Direction-specific fixed windows $d_F^x$ and $d_F^y$ may be introduced to represent roller or pinned boundary conditions. The process of boundary condition imposition is visualised in \FIG{fig:displacement_field_enforcement}.

\paragraph{Phase field}

The physical phase-field variable satisfies the constraint $0 \le \bar{c} \le 1$. The NN output $c$ is unconstrained and is therefore transformed to enforce admissibility. Sigmoid-based mappings may suffer from vanishing gradient effects in saturated regions. A piecewise transformation inspired by \cite{Manav2024} is adopted. The constrained phase-field variable is defined as
\begin{equation}
\label{eqn:pf_mag}
\begin{aligned}
\bar{c} = f_c(c) =
\begin{cases}
\beta(c+\ell_{s}), & c < -\ell_{s}, \\
\dfrac{c}{2\ell_{s}} + \dfrac{1}{2}, & |c| \le \ell_{s}, \\
\beta(c-\ell_{s}), & c > \ell_{s}.
\end{cases}
\end{aligned}
\end{equation}

This mapping preserves a linear regime within $[-\ell_{s},\ell_{s}]$ and penalises excursions outside this interval while avoiding strong saturation effects.

\subsection{Numerical quadrature}
\label{sec:numeric_quadrature_rule}

Although PINNs are often formulated without structured meshes, the present approach employs a finite element-style spatial discretisation for energy evaluation. This enables a well-established and computationally efficient framework for evaluating spatial derivatives. It also permits direct comparison with classical finite element solutions for validation purposes. In the context of variational fracture modelling, the use of automatic differentiation for higher-order derivatives has been reported to induce oscillatory behaviour in the displacement field,\cite{Manav2024} which may affect stress evaluation near crack tips.

The anisotropic phase-field formulation considered herein involves higher-order gradient terms. Such formulations require $C^1$ continuity across element boundaries. Uniform B-spline meshes are therefore adopted to ensure the required inter-element smoothness and consistent evaluation of second-order spatial derivatives. In the benchmark tests presented in this work, numerical integration is performed using a $3 \times 3$ Legendre-Gauss quadrature rule, corresponding to nine Gauss points per element for two-dimensional problems.

\subsection{Optimisation algorithms}\label{sec:optimisation_algorithm}

Recent developments in PINN methodologies have demonstrated improved convergence behaviour for the solution of partial differential equations.\cite{Kiyani2025} In fracture mechanics applications, combinations of first-order and second-order optimisation methods have been reported to provide favourable performance.\cite{Goswami2020a, Manav2024}

The phase-field constraint function shown in \FIG{fig:nonsmooth_sigmoid} implies that, under typical random weight initialisation centred around zero, the initial phase-field prediction is approximately $0.5$. To recover the undamaged state prior to loading, a pre-training step is introduced using a near-zero displacement value. For the benchmarks considered in this work, $U_0 = \num{1e-12}~\mathrm{m}$ is adopted. This stabilises the initial optimisation and ensures that the phase field converges to the intact configuration before incremental loading is applied.

Second-order optimisation methods exhibit rapid convergence during the pre-training phase. The limited-memory Broyden-Fletcher-Goldfarb-Shanno (LBFGS) algorithm is commonly adopted in the literature.\cite{Goswami2020a, Manav2024} In the present work, self-scaled Broyden family methods, recently reintroduced to the PINN literature by Kiyani \emph{et al.}\cite{Kiyani2025} are employed. The self-scaled Broyden method with Armijo backtracking demonstrates robust convergence within several hundred epochs. For sufficiently large network architectures, storage of the full dense history matrix becomes computationally prohibitive. A truncated LBFGS-like history strategy is therefore adopted.\cite{Byrd_Nocedal_Schnabel_1994}

Following pre-training, the applied displacement is introduced incrementally. The displacement increments must be sufficiently small to avoid large energy barriers between successive states. Excessively small increments, however, may amplify accumulated optimisation errors arising from incomplete convergence at intermediate steps.

Three optimisation strategies are examined: (a) combined first-order and second-order optimisation (Adam\cite{Kingma_Ba_2014} followed by LBFGS or Broyden with Armijo line search), (b) second-order optimisation only (Broyden with Armijo), and (c) first-order optimisation only (Resilient Backpropagation - RPROP). The combined strategy (a) achieves convergence when displacement is applied instantaneously. However, reduced stability is observed under incremental loading. This behaviour is consistent with the increased non-convexity introduced by irreversibility constraints. The second-order strategy (b) produces correct crack paths under incremental loading. In certain cases, however, the crack growth rate exhibits irregular progression compared to finite element reference solutions. The first-order strategy (c), previously applied to fracture mechanics by Manav \emph{et al.}\cite{Manav2024} yields smooth crack evolution and respects energy barriers during incremental loading. Convergence is slower than with second-order methods, but the resulting crack growth behaviour is consistent with finite element simulations. Accordingly, strategy (c) is adopted for all benchmark results presented in this work.

\subsubsection{Transfer learning}\label{sec:transfer_learning}

The phase-field solution at each load increment depends on the previously obtained state. Transfer learning is therefore a natural strategy within the incremental loading framework.\cite{Goswami2020a} At each increment, the network parameters from the converged previous step are used to initialise the optimisation. These parameters include the weights $\mathbf{w}$, biases $\mathbf{b}$, activation scaling coefficient $r$, and residual scaling coefficients $\alpha$.

For sufficiently small displacement increments, only limited changes in the energy functional are introduced between successive steps. As a result, the network parameters remain close to a local minimum of the updated loss functional. This significantly reduces the number of iterations required for convergence, which remains below $500$ optimisation steps per increment for the benchmarks examined in this study.

For first-order optimisation schemes such as RPROP, the adaptive step sizes obtained at the previous increment are also transferred. This reduces the need for re-adjustment of the learning dynamics at each load step. In contrast, transferring curvature information in second-order or quasi-Newton schemes may adversely affect convergence. Reusing history matrices from previous increments can bias the search direction toward previously identified minima, potentially limiting exploration of newly introduced non-convex regions of the loss landscape.

\subsubsection{Early stopping}\label{sec:early_stopping}

Convergence of the optimisation procedure is assessed using a relative loss criterion. Training is terminated when the relative change in the loss functional remains below a prescribed tolerance over a specified number of consecutive epochs. This approach follows established practice in recent phase-field PINN formulations.\cite{Manav2024} A relative criterion is adopted in preference to an absolute threshold to account for the evolving magnitude of the loss functional during incremental loading. This prevents premature termination at early load steps and avoids excessive optimisation effort at later stages.

\section{Numerical results} \label{sec:num_results}

This section demonstrates the capability of the proposed model using the standard square plate under pure tension benchmark (Section \ref{sec:sq_plate_tension}). Results obtained with the proposed DRM formulation are compared against reference FEM solutions. The FEM simulations were performed on an Intel i9-14900K CPU using 24 cores, while the DRM simulations were carried out on NVIDIA A100-PCIE-40GB GPUs. Simulation timings are reported in each section. The NN hyperparameters used in the simulations are summarised in \TAB{tab:nn_hyperparams}.

\begin{table}[htbp]
    \centering
    \caption{NN architecture and training hyperparameters.}
    \label{tab:nn_hyperparams}
    \begin{tabular}{ll}
        \toprule
        \textbf{Parameter} & \textbf{Value} \\
        \midrule
        \multicolumn{2}{l}{\textit{Architecture}} \\
        \midrule
        Blocks / Depth $\mathcal{D}$ / Width                  & $6$ / $4$ / $300$ \\
        RFF feature width $m$                      & $192$ \\
        Standard deviation $\sigma_{\mathrm{rff}}$                 & $0.1$ \\
        Initial activation scaling $r$                  & $2$ \\
        Weight initialisation                   & Glorot normal \\
        Bias initialisation                     & $0$ \\
        Residual scaling $\alpha$                  & $0$ \\
        \midrule
        \multicolumn{2}{l}{\textit{Phase-field constraints}} \\
        \midrule
        Out-of-bound slope $\beta$                                 & $10^{-3}$ \\
        Main bound length $\ell_{s}$                              & $8.0$ \\
        Irreversibility coefficient $\kappa_{\mathrm{irr}}$                 & $0.02$ \\
        Residual stiffness $k_f$                & $0$ \\
        \midrule
        \multicolumn{2}{l}{\textit{Training}} \\
        \midrule
        Weight regularisation coefficient $\kappa_{\mathrm{reg}}$                 & $10^{-5}$ \\
        Early stopping criterion                & $\Delta\mathcal{L}_{\mathrm{rel}} < 5\times10^{-5}$ over 50 epochs \\
        Pre-training epochs (max)               & $1000$ \\
        Pre-training optimiser                  & Broyden (Armijo backtracking, memory $= 10$) \\
        Training epochs (max)               & $10000$ \\
        Training optimiser                & RPROP (default parameters) \\
        Random seeds                            & $1$--$10$ \\
        \bottomrule
    \end{tabular}
\end{table}

The pseudo-random number generator is initialised at the start of each simulation, with seeds ranging from $1$ to $10$ to assess sensitivity to random initialisation. The reported energy plots show the mean and standard deviation across all seed values. The code used to generate the FEM and DRM results, together with the data presented in this section, is available through the associated Zenodo repository \cite{plunge_2026_19553764}.

\subsection{Square plate under pure tension} \label{sec:sq_plate_tension}

A square plate subjected to pure tension is analysed to assess the capability of the proposed framework to capture crack nucleation and propagation under different anisotropic fracture conditions. Two groups of scenarios are considered.

In the first group, three material symmetry cases are examined: isotropic behaviour, cubic anisotropy, and orthotropic anisotropy. For the anisotropic cases, the relationship between crack propagation direction and material orientation is investigated. This allows the influence of material symmetry on the resulting fracture paths to be examined.

In the second group, a layered plate with alternating material orientations is considered. Each layer follows an orthotropic symmetry. The objective of this configuration is to examine the interaction between crack propagation and abrupt changes in material orientation, and to assess the ability of the proposed framework to capture crack deflection across material interfaces.

In the third group, convergence analysis is performed on the orthotropic anisotropy case with a material orientation of $30^{\circ}$. The influence of displacement increment size and mesh resolution on the predicted crack path and energy evolution is investigated. 

\begin{figure}[htbp]
    \centering
    \begin{tabular}{c@{\hspace{6em}}c}
    \subfloat[\label{fig:sq_plate_simple}]{\includegraphics[width=0.3\textwidth]{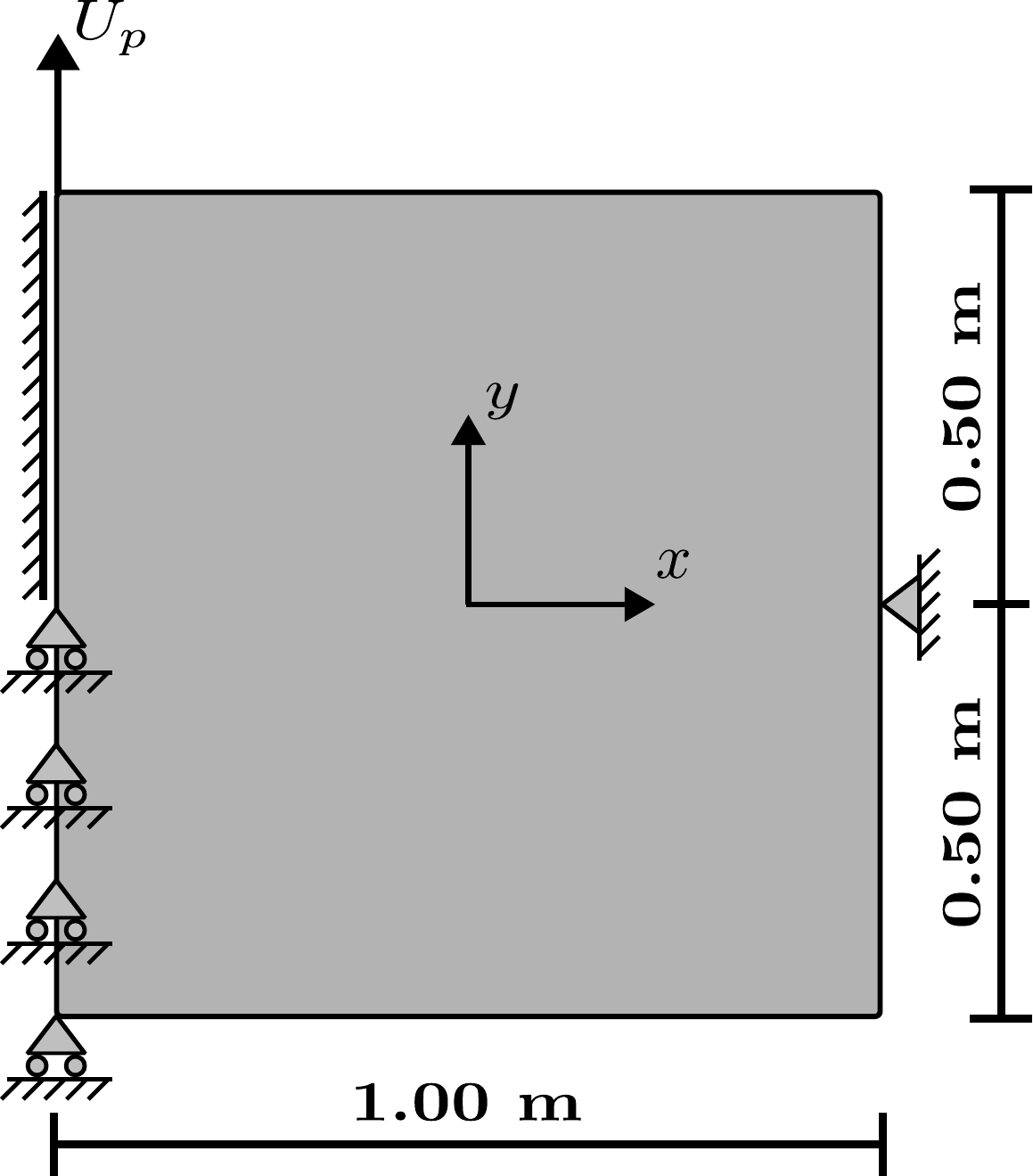}} &
    \subfloat[\label{fig:sq_plate_alternating}]{\includegraphics[width=0.3\textwidth]{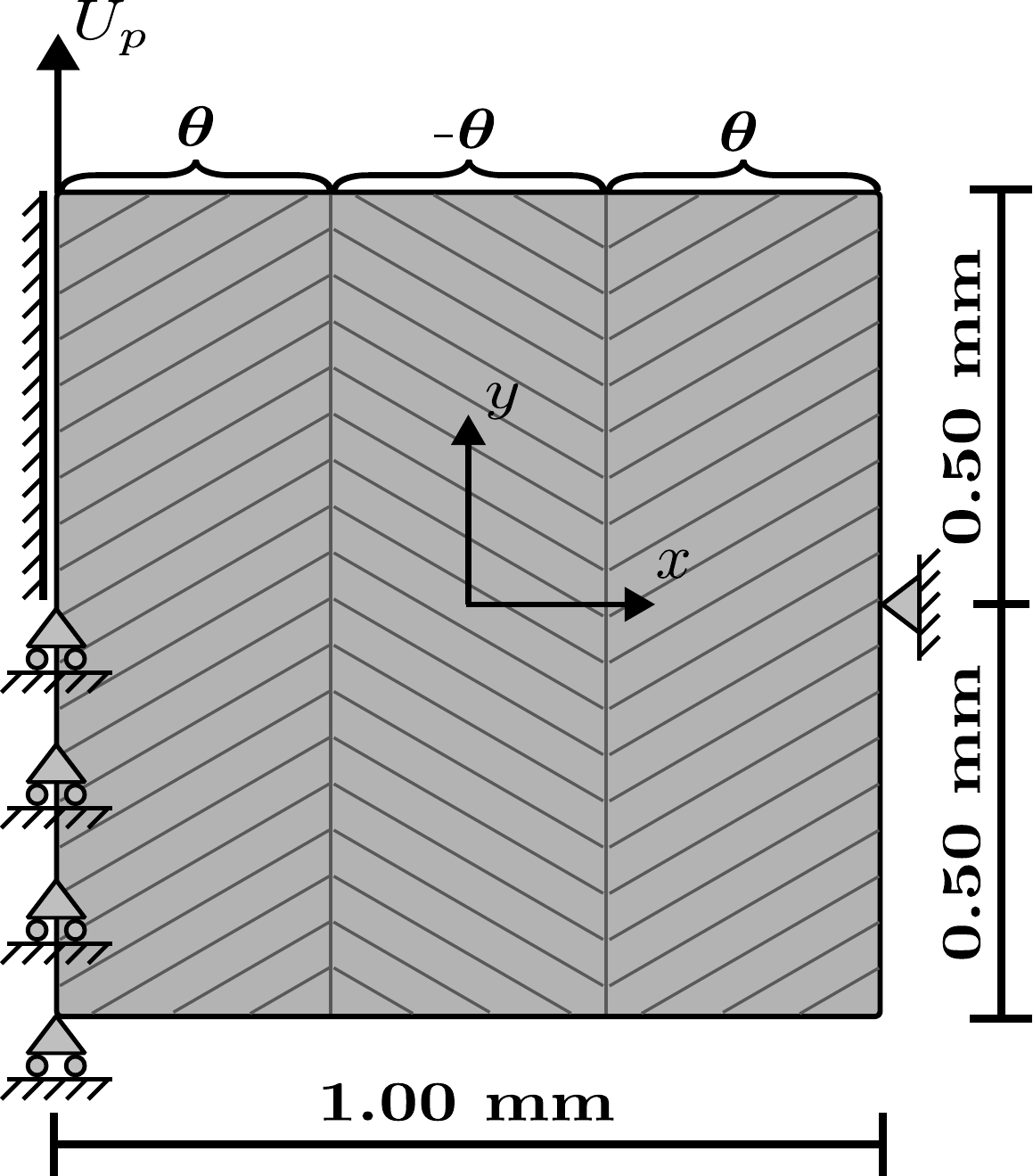}}
    \end{tabular}
    \caption[]{Square plate under pure tension: geometry and boundary conditions: \subref{fig:sq_plate_simple} homogeneous plate with uniform material properties, \subref{fig:sq_plate_alternating} layered plate with alternating material orientations.}
    \label{fig:sq_plate_tension_geo}
\end{figure}

The geometry and boundary conditions of the problem are shown in \FIG{fig:sq_plate_tension_geo}. The lower half of the left edge is constrained in the $y$-direction, with a pin support at the midpoint of the right edge restraining both $x$- and $y$-displacements. The plate is loaded under displacement control by prescribing vertical displacements along the upper half of the left edge. Plane strain conditions are assumed. The phase-field length-scale parameter is set to $l_0 = 0.01$ m.

Results from all benchmarks were compared against their FEM counterparts to assess the accuracy of the proposed method. In particular, the evolution of elastic strain and fracture energies over the displacement increments was examined, together with the corresponding phase-field distributions. The results consistently overpredict the elastic energy, with the discrepancy becoming more pronounced as crack initiation and propagation develop. This behaviour is attributed to the use of smooth activation functions in the approximation of both the displacement and phase fields, which may limit the ability of the model to represent the displacement discontinuity associated with fracture. This observation is consistent with the findings of Manav \emph{et al.},\cite{Manav2024} who reported improved agreement in the energy response when employing ReLU activation functions instead of $\tanh$.

\paragraph{Material properties and anisotropy parameters}

\begin{figure}[htbp]
	\centering
    \begin{tabular}{c@{\hspace{4em}}c}
		\subfloat[\label{fig:gc_theta_params:a}]{
        \includegraphics[width=0.35\textwidth]{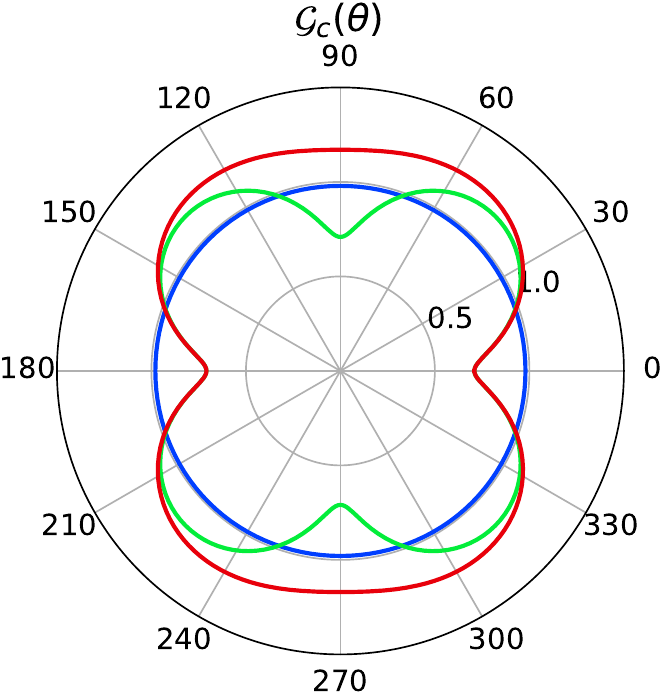}} &
		\subfloat[\label{fig:gc_theta_params:b}]{
			\includegraphics[width=0.35\textwidth]{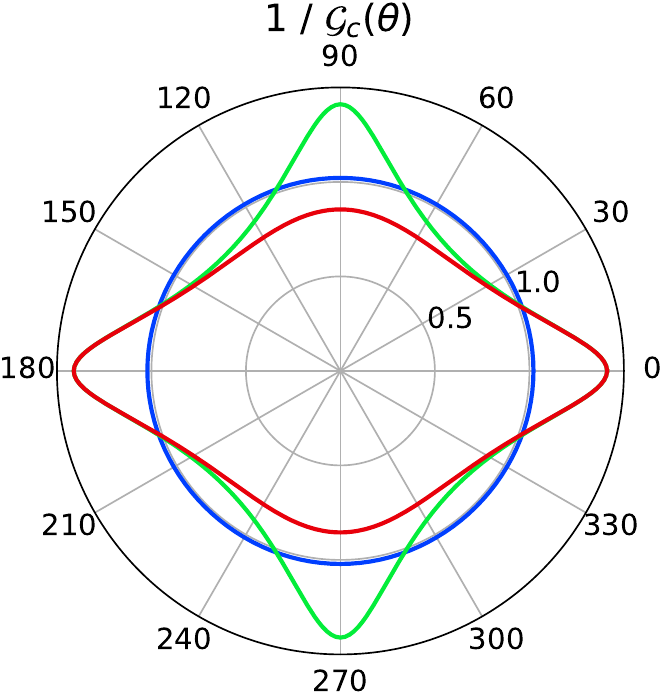}}
	\end{tabular}
\caption[]{\subref{fig:gc_theta_params:a} Polar representation of the fracture surface energy density $\mathcal{G}_{c}(\theta)$ and \subref{fig:gc_theta_params:b} its reciprocal $1/\mathcal{G}_{c}(\theta)$ for the different material symmetries as functions of crack orientation $\theta$. The blue, green, red lines are corresponding to the isotropic, cubic and orthotropic cases, respectively.}
    \label{fig:gc_theta_params}
\end{figure}

Three material symmetry cases are considered. The elastic behaviour is described using linear elasticity, while anisotropic fracture behaviour is introduced through the tensorial coefficients $\gamma_{ijkl}$ in the crack density functional. The constitutive matrices are provided in Appendix~\ref{app:constitutive}. The polar plots for the fracture anisotropy parameters are presented in \FIG{fig:gc_theta_params}.

\begin{itemize}
\item \emph{Isotropic material:} Young’s modulus: $E = 1~\text{GPa}$. Poisson’s ratio: $\nu = 0.30$. Fracture anisotropy parameters: $\gamma_{1111}=\gamma_{2222}=\gamma_{1122}=\gamma_{1212}=0$.
\item \emph{Cubic material (in material principal axis):} Young's modulus: $E = 1~\text{GPa}$. Poisson's ratio: $\nu = 0.3$. Shear modulus is set to be half of the isotropic form: $G = 0.5E/(2(1+\nu))\approx 0.192~\text{GPa}$. Fracture anisotropy parameters: $\gamma_{1111}=1$, $\gamma_{2222}=1$, $\gamma_{1122}=0$ and $\gamma_{1212}=74$.
\item \emph{Orthotropic material (in material principal axis):} Young's moduli: $E_{11}=20$~GPa, $E_{22}=E_{33}=1$~GPa. Poisson's ratios: $\nu_{12}=0.30$, $\nu_{13}=0.30$, $\nu_{23}=0.35$. Shear moduli are computed using the isotropic relation applied to the more compliant direction in each plane: $G_{ij} = \min(E_{ii},\,E_{jj})/(2(1+\nu_{ij}))$, giving $G_{12}=E_{22}/(2(1+\nu_{12}))\approx 0.385$~GPa, $G_{13}=E_{33}/(2(1+\nu_{13}))\approx 0.385$~GPa and $G_{23}=E_{22}/(2(1+\nu_{23}))\approx 0.370$~GPa. Fracture anisotropy parameters: $\gamma_{1111}=20$, $\gamma_{2222}=1$, $\gamma_{1122}=0$ and $\gamma_{1212}=74$.

\end{itemize}

\subsubsection{Crack propagation under different material symmetries} \label{sec:sq_plate_tension_mat_orient}

The first set of simulations investigates crack propagation in a homogeneous plate for different material symmetry conditions. Five cases are considered: an isotropic material, two cubic anisotropy cases with material orientations of $-30^{\circ}$ and $-50^{\circ}$, and two orthotropic anisotropy cases with the same orientations. The objective is to assess the capability of the proposed framework to capture the influence of material symmetry and orientation on fracture behaviour.

For each configuration, the evolution of elastic and fracture energies is analysed as a function of the applied displacement. The predicted phase-field distributions obtained with the proposed DRM are compared with reference solutions from the FEM considering the same mesh size, i.e. $1/201$ m. FEM simulations were performed over $500$ displacement increments, while DRM simulations used $200$ increments, both spanning $[0,\, U_{\text{max}}]$ m. Each case is illustrated using three subfigures: the energy evolution, the FEM phase-field distribution, and the DRM phase-field distribution.

\begin{figure}[htbp]
    \centering
    \begin{tabular}{>{\centering\arraybackslash}m{0.45\textwidth} 
            >{\centering\arraybackslash}m{0.25\textwidth} 
        >{\centering\arraybackslash}m{0.25\textwidth}}
        \subfloat[\label{fig:isotropic_case:a}]{
        \includegraphics[width=0.45\textwidth]{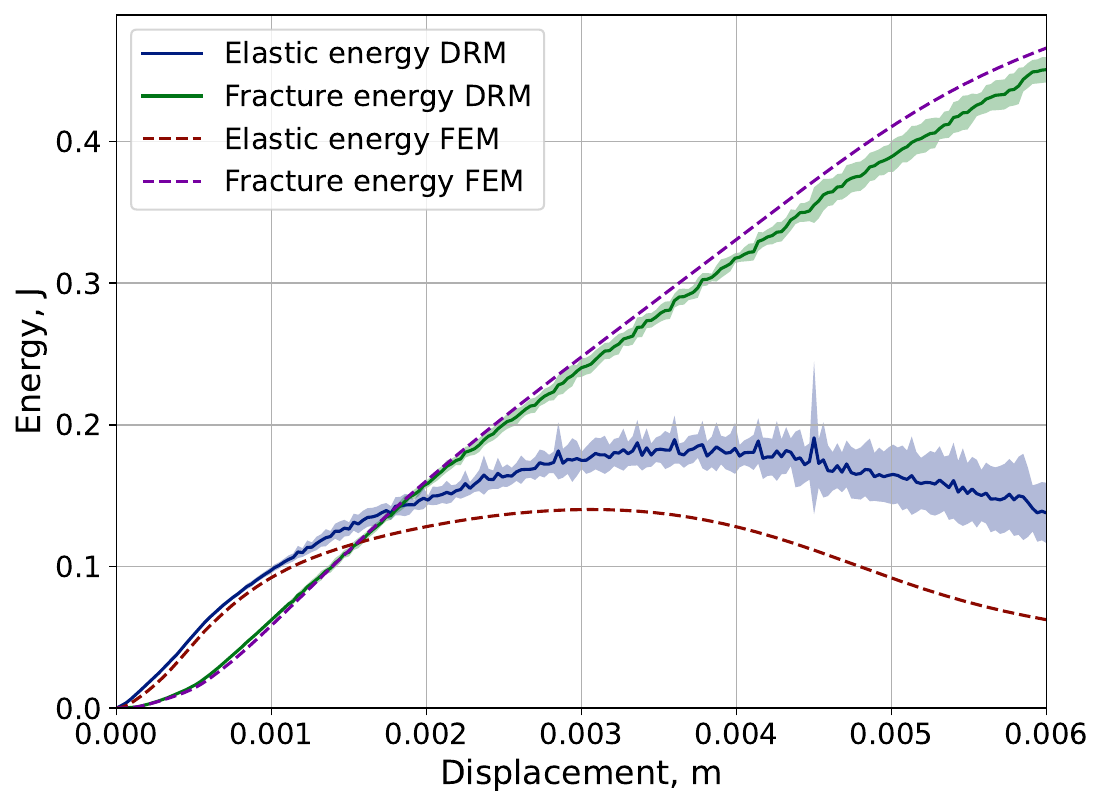}} &
        \subfloat[\label{fig:isotropic_case:b}]{
    \includegraphics[width=0.25\textwidth]{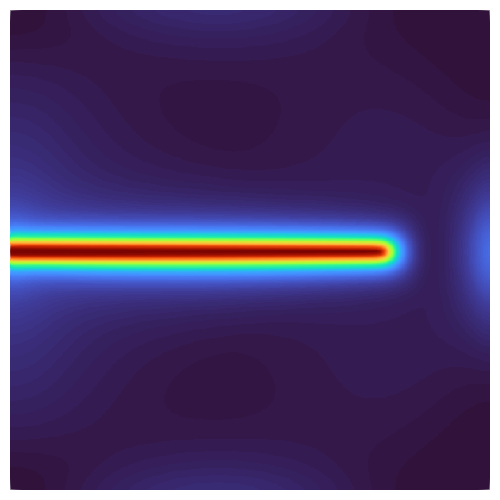}} &
    \subfloat[\label{fig:isotropic_case:c}]{
    \includegraphics[width=0.25\textwidth]{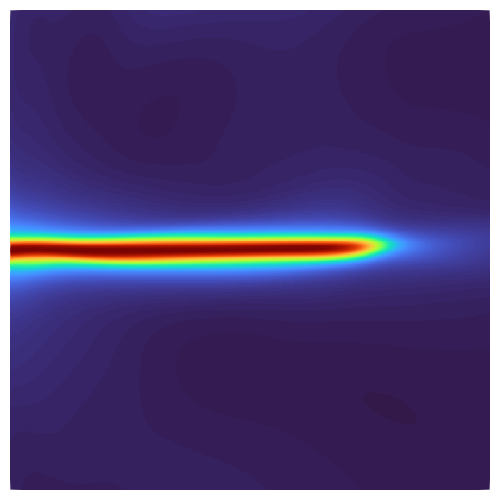}} \\
    \multicolumn{3}{c}{\includegraphics[width=0.25\textwidth]{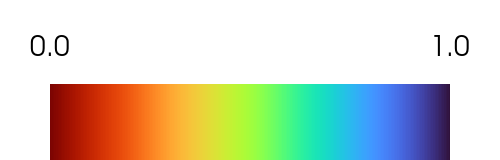}}
\end{tabular}
\caption[]{Results for the isotropic symmetry case. 
    \subref{fig:isotropic_case:a} Evolution of elastic and fracture energies as a function of displacement. 
    \subref{fig:isotropic_case:b} Phase-field distribution obtained using the FEM. 
\subref{fig:isotropic_case:c} Phase-field distribution obtained using the proposed DRM at a displacement of $0.006~\text{m}$. The fracture trajectory is generated by seed number $2$.}
\label{fig:isotropic_case}
\end{figure}

The isotropic case is presented in \FIG{fig:isotropic_case}. The energy evolution shown in \FIG{fig:isotropic_case:a} exhibits the characteristic behaviour of brittle fracture. The elastic energy increases with the applied displacement until crack initiation occurs. After crack nucleation, fracture energy increases while the elastic energy decreases due to stress redistribution. The FEM phase-field distribution in \FIG{fig:isotropic_case:b} shows a straight crack path propagating perpendicular to the loading direction. The DRM prediction in \FIG{fig:isotropic_case:c} reproduces the same crack trajectory and energy evolution. The close agreement between the two approaches confirms the accuracy of the proposed framework for isotropic fracture. The DRM solutions converged in $303\pm37$ (mean $\pm$ std) minutes across the $10$ random seeds.

\begin{figure}[htbp]
	\centering
    \begin{tabular}{>{\centering\arraybackslash}m{0.45\textwidth} 
            >{\centering\arraybackslash}m{0.25\textwidth} 
        >{\centering\arraybackslash}m{0.25\textwidth}}
        \subfloat[\label{fig:cubic_case_30:a}]{
            \includegraphics[width=0.45\textwidth]{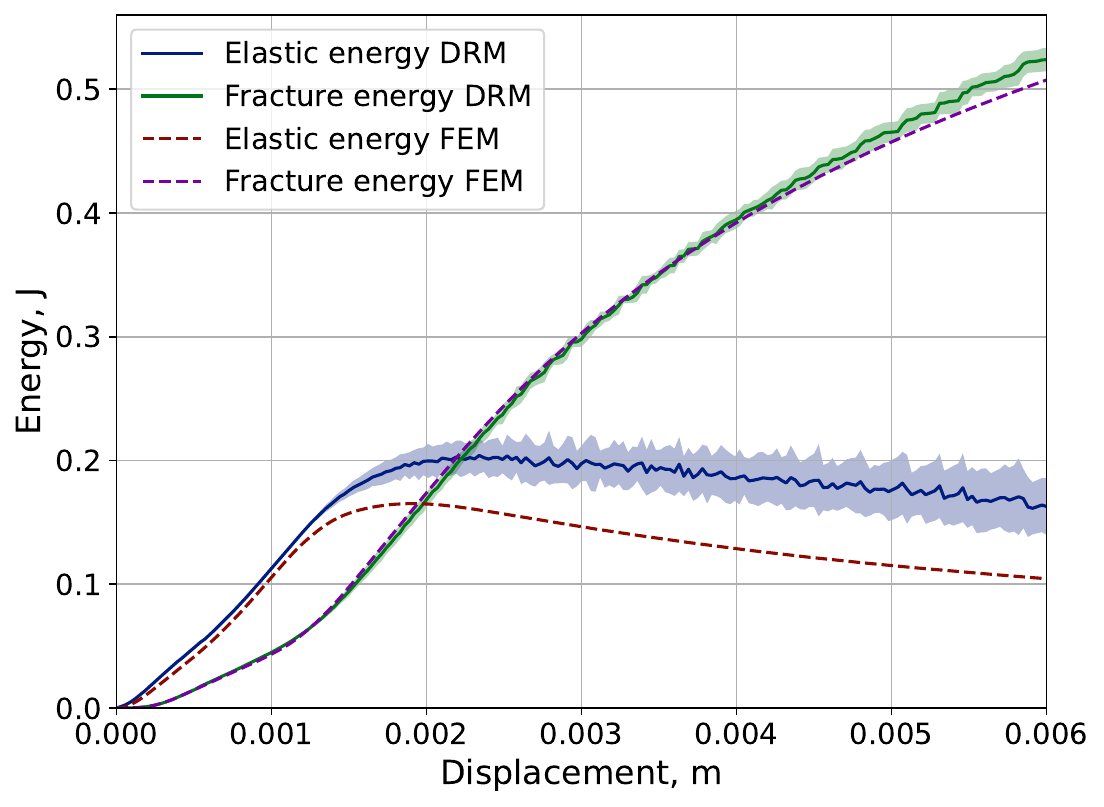}} &
        \subfloat[\label{fig:cubic_case_30:b}]{
            \includegraphics[width=0.25\textwidth]{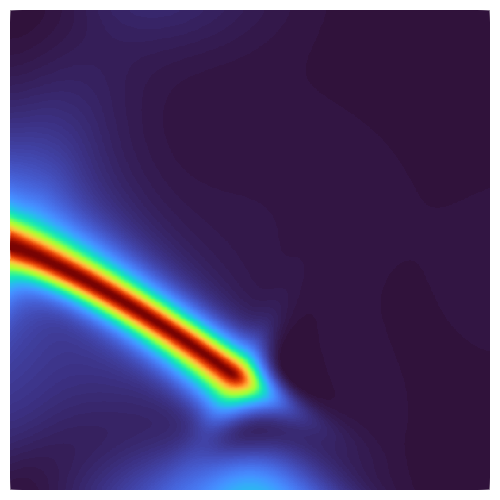}} &
        \subfloat[\label{fig:cubic_case_30:c}]{
            \includegraphics[width=0.25\textwidth]{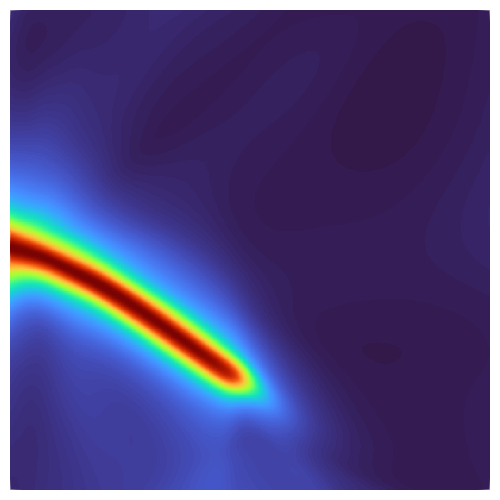}} \\
        \subfloat[\label{fig:cubic_case_50:d}]{
            \includegraphics[width=0.45\textwidth]{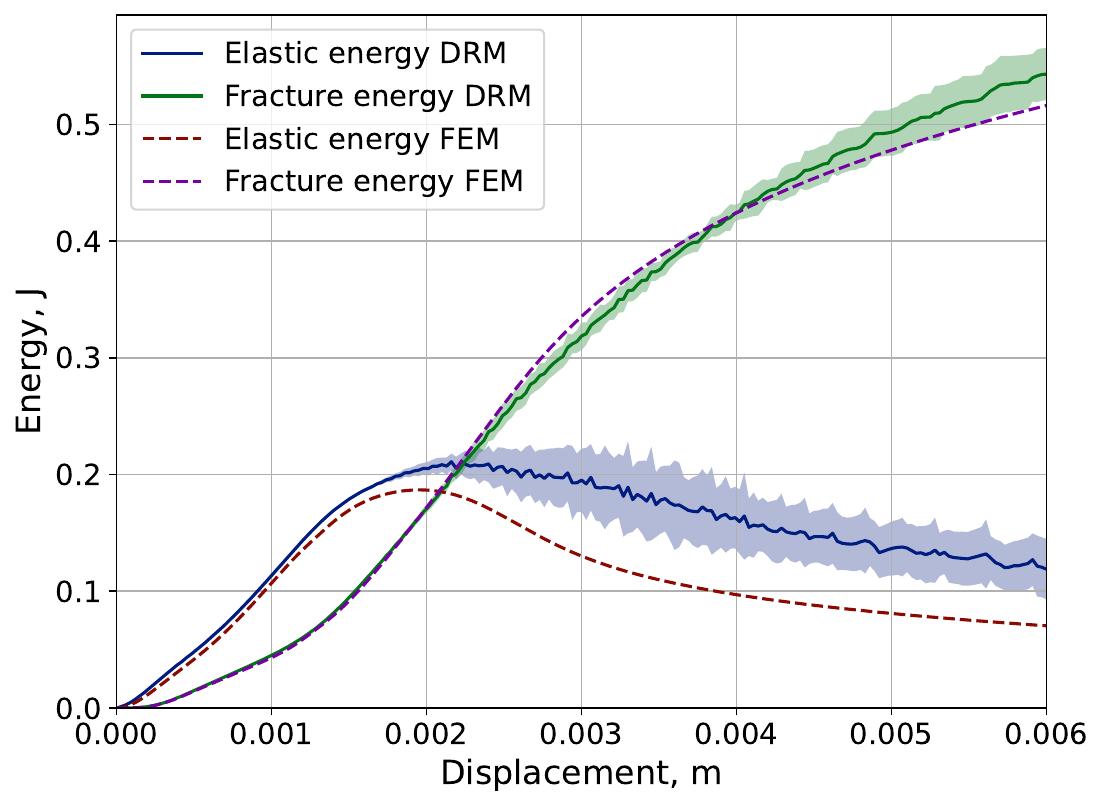}} &
        \subfloat[\label{fig:cubic_case_50:e}]{
            \includegraphics[width=0.25\textwidth]{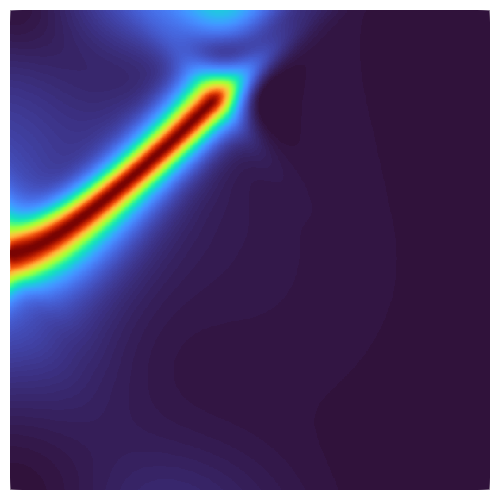}} &
        \subfloat[\label{fig:cubic_case_50:f}]{
            \includegraphics[width=0.25\textwidth]{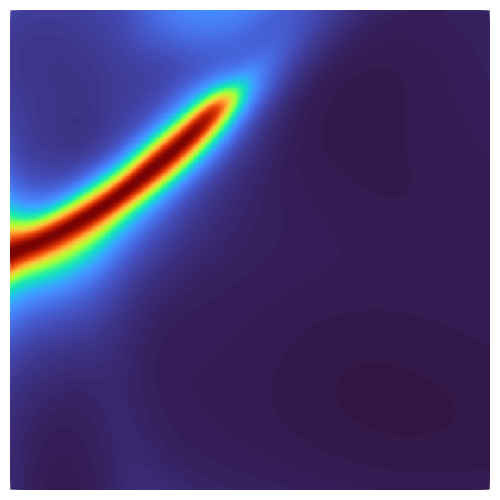}} \\
        \multicolumn{3}{c}{\includegraphics[width=0.25\textwidth]{colorbar.png}}
	\end{tabular}
	\caption[]{Cubic material symmetry for two material orientations. 
    Material orientation $-30^{\circ}$: \subref{fig:cubic_case_30:a} evolution of elastic and fracture energies as a function of displacement,  \subref{fig:cubic_case_30:b} FEM phase-field distribution, and  \subref{fig:cubic_case_30:c} DRM phase-field distribution at a displacement of $0.006~\text{m}$. The fracture trajectory is generated by seed number $10$. Material orientation $-50^{\circ}$: \subref{fig:cubic_case_50:d} evolution of elastic and fracture energies as a function of displacement,  \subref{fig:cubic_case_50:e} FEM phase-field distribution, and  \subref{fig:cubic_case_50:f} DRM phase-field distribution at a displacement of $0.006~\text{m}$. The fracture trajectory is generated by seed number $7$.}
    \label{fig:cubic_ortho_case_30}
\end{figure}

The influence of cubic anisotropy is illustrated in \FIG{fig:cubic_ortho_case_30}. The first row corresponds to a material orientation of $-30^{\circ}$, shown in \FIG{fig:cubic_case_30:a}--\FIG{fig:cubic_case_30:c}. The crack path deviates from the vertical direction and progressively aligns with the preferred fracture direction defined by the cubic anisotropy. When the orientation increases to $-50^{\circ}$, shown in \FIG{fig:cubic_case_50:d}--\FIG{fig:cubic_case_50:f}, the deviation becomes more pronounced. The crack reorients towards the energetically favourable material direction. In both cases, the DRM phase-field distributions closely match the FEM solutions and reproduce the corresponding energy evolution. For material orientation $-30^{\circ}$, seed 8 failed to converge; for $-50^{\circ}$, seeds 5 and 6 failed. These runs are excluded from the aggregate energy plots. Solution times for the DRM were $285\pm30$ and $281\pm26$ minutes (mean $\pm$ std, 10 random seeds) for $-30^{\circ}$ and $-50^{\circ}$, respectively.

\begin{figure}[htbp]
	\centering
    \begin{tabular}{>{\centering\arraybackslash}m{0.45\textwidth} 
            >{\centering\arraybackslash}m{0.25\textwidth} 
        >{\centering\arraybackslash}m{0.25\textwidth}}
        \subfloat[\label{fig:ortho_case_30:a}]{
            \includegraphics[width=0.45\textwidth]{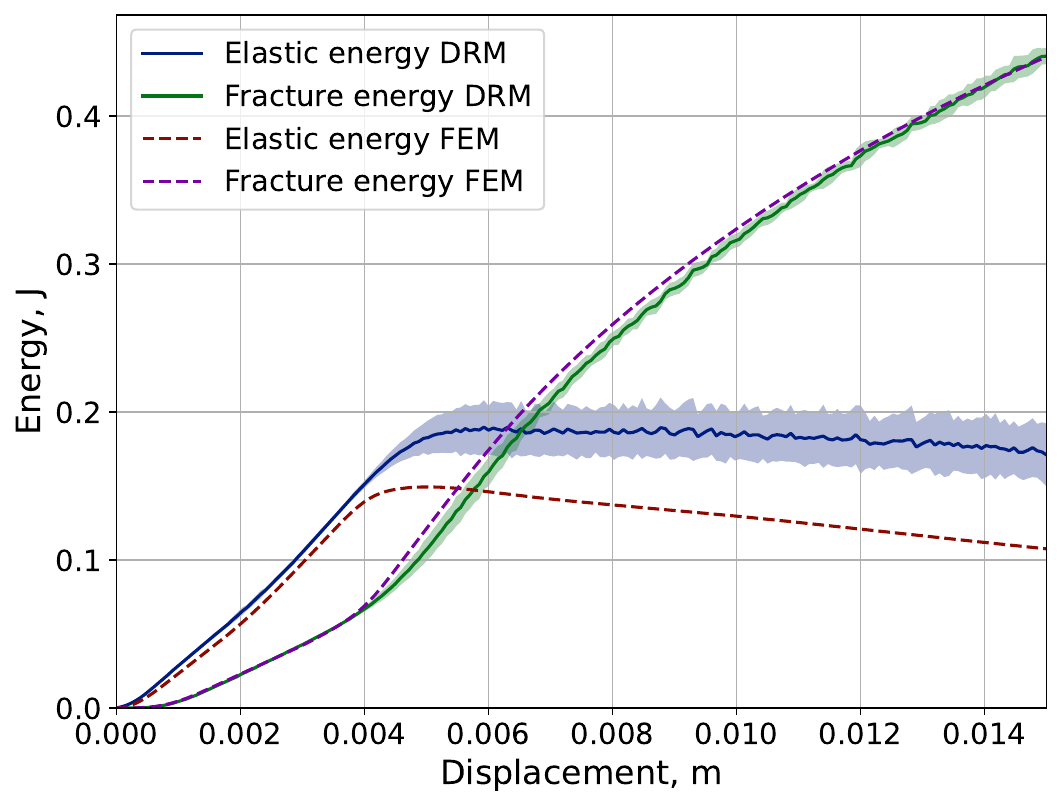}} &
        \subfloat[\label{fig:ortho_case_30:b}]{
            \includegraphics[width=0.25\textwidth]{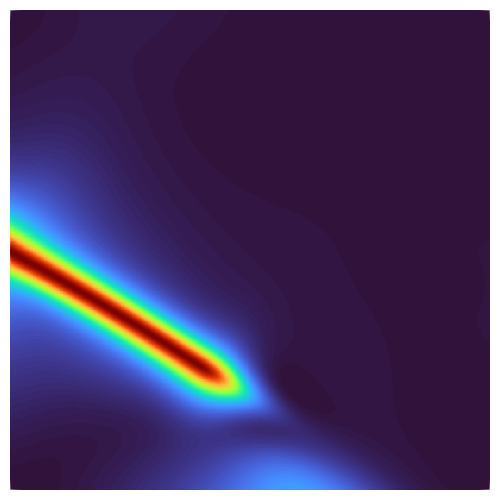}} &
        \subfloat[\label{fig:ortho_case_30:c}]{
            \includegraphics[width=0.25\textwidth]{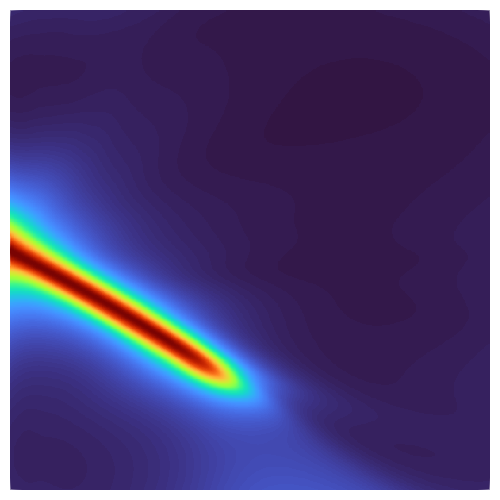}} \\
        \subfloat[\label{fig:ortho_case_50:d}]{
            \includegraphics[width=0.45\textwidth]{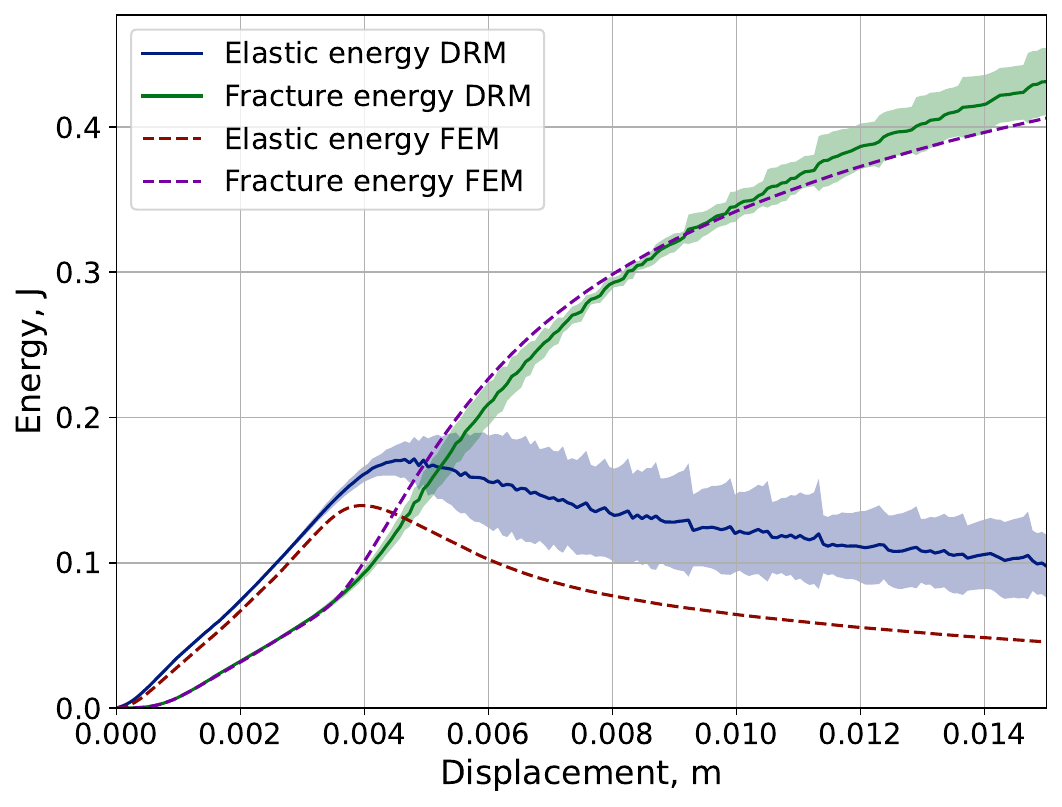}} &
        \subfloat[\label{fig:ortho_case_50:e}]{
            \includegraphics[width=0.25\textwidth]{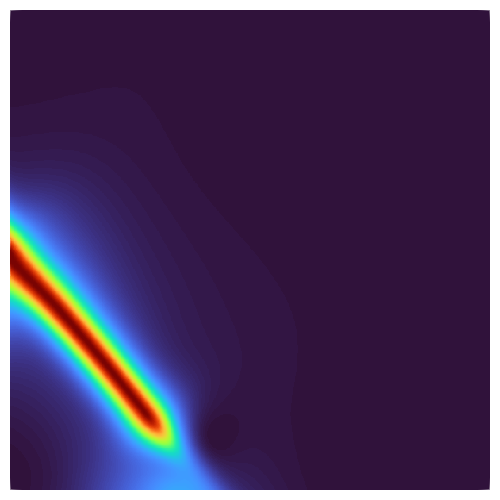}} &
        \subfloat[\label{fig:ortho_case_50:f}]{
            \includegraphics[width=0.25\textwidth]{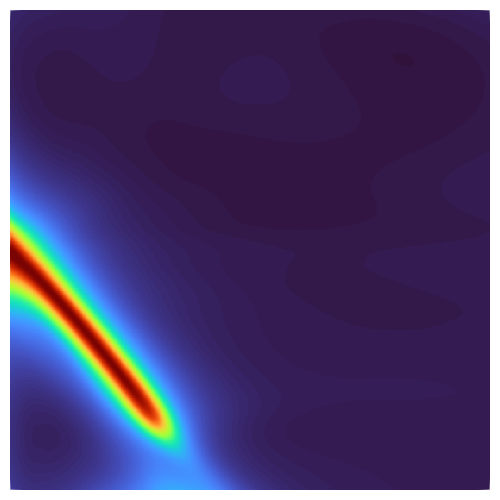}} \\
        \multicolumn{3}{c}{\includegraphics[width=0.25\textwidth]{colorbar.png}}
	\end{tabular}
    \caption[]{Orthotropic material symmetry for two material orientations. Material orientation $-30^{\circ}$:  \subref{fig:ortho_case_30:a} variation of elastic and fracture energies with applied displacement,  \subref{fig:ortho_case_30:b} phase-field distribution obtained using the finite element method (FEM), and \subref{fig:ortho_case_30:c} phase-field distribution obtained using the proposed DRM at a displacement of $0.015~\text{m}$. The fracture trajectory is generated by seed number $1$. Material orientation $-50^{\circ}$:  \subref{fig:ortho_case_50:d} corresponding energy evolution,  \subref{fig:ortho_case_50:e} FEM phase-field distribution, and  \subref{fig:ortho_case_50:f} DRM phase-field distribution at a displacement of $0.015~\text{m}$. The fracture trajectory is generated by seed number $3$.}
    \label{fig:ortho_case_50}
\end{figure}

The orthotropic cases are presented in \FIG{fig:ortho_case_50}. For the $-30^{\circ}$ orientation, the results are shown in \FIG{fig:ortho_case_30:a}--\FIG{fig:ortho_case_30:c}. The stronger directional dependence of the orthotropic fracture parameters produces a more pronounced deviation of the crack path compared with the cubic case. The crack rapidly aligns with the preferred material direction. For the $-50^{\circ}$ orientation, shown in \FIG{fig:ortho_case_50:d}--\FIG{fig:ortho_case_50:f}, the anisotropy further accentuates this behaviour. The crack path is strongly controlled by the orthotropic symmetry and propagates along the dominant material direction. The DRM predictions again show very good agreement with the FEM results in both crack trajectory and energy evolution. Solution times for the DRM were $294\pm41$ and $300\pm45$ minutes (mean $\pm$ std, $10$ random seeds) for $-30^{\circ}$ and $-50^{\circ}$, respectively. In comparison, each FEM simulation required approximately $60$ minutes to converge.

\begin{figure}[htbp]
	\centering
    \begin{tabular}{>{\centering\arraybackslash}m{0.45\textwidth} 
            >{\centering\arraybackslash}m{0.25\textwidth} 
        >{\centering\arraybackslash}m{0.25\textwidth}}
        \subfloat[\label{fig:manav_comparison:a}]{
            \includegraphics[width=0.45\textwidth]{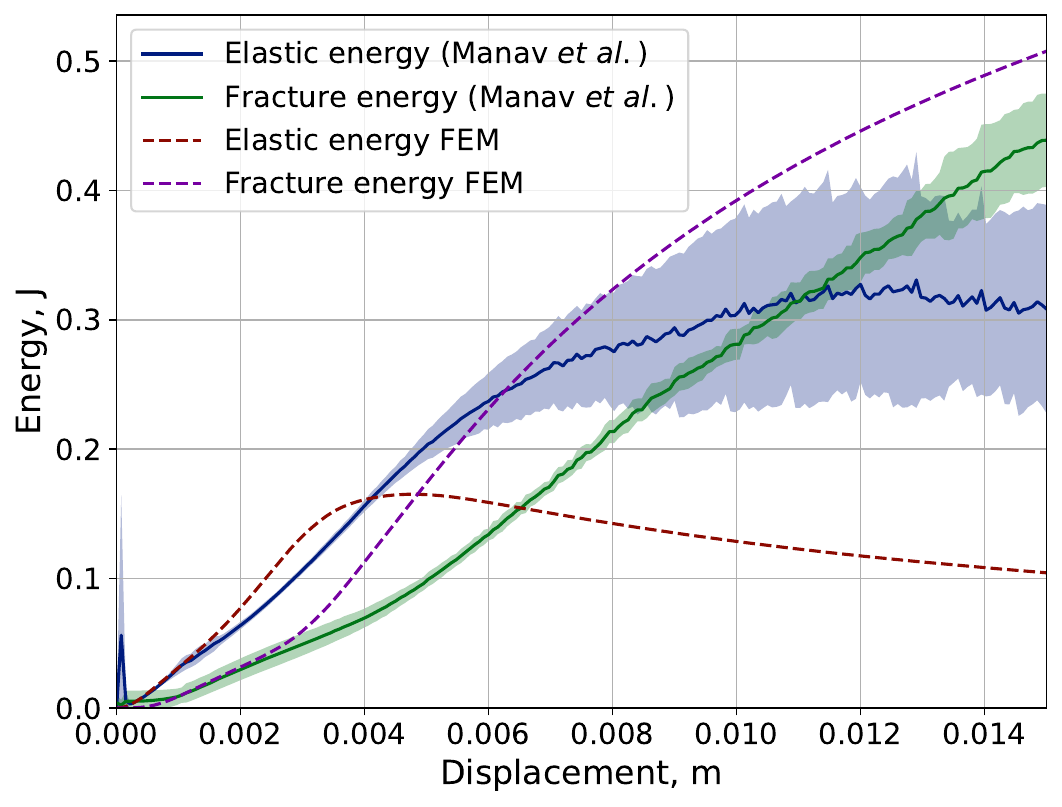}} &
        \subfloat[\label{fig:manav_comparison:b}]{
            \includegraphics[width=0.25\textwidth]{fem_oa+30.png}} &
        \subfloat[\label{fig:manav_comparison:c}]{
            \includegraphics[width=0.25\textwidth]{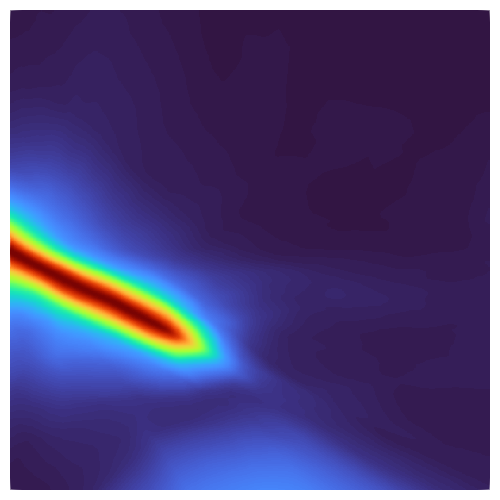}} \\
        \multicolumn{3}{c}{\includegraphics[width=0.25\textwidth]{colorbar.png}}
	\end{tabular}
    \caption[]{Orthotropic material symmetry of $-30^{\circ}$ tested with network architecture introduced by Manav \emph{et al.} for second-order fracture problems:  \subref{fig:manav_comparison:a} variation of elastic and fracture energies with applied displacement,  \subref{fig:manav_comparison:b} phase-field distribution obtained using the finite element method (FEM), and \subref{fig:manav_comparison:c} phase-field distribution obtained using the proposed DRM at a displacement of $0.015~\text{m}$. The fracture trajectory is generated by seed number $3$.}
    \label{fig:manav_comparison}
\end{figure}

To demonstrate the necessity of the architectural modifications proposed in this work, the orthotropic case with a material orientation of $-30^{\circ}$ is simulated using the hyperparameters adopted by Manav \emph{et al.} Specifically, a fully-connected network comprising 8 layers, each 400 neurons wide, is employed. The initial coefficient for the trainable ReLU activation function is set to $3$, and the early stopping criterion is tightened to a relative loss tolerance of $\Delta\mathcal{L}_{\mathrm{rel}} < 5\times10^{-6}$. The resulting predictions are presented in \FIG{fig:manav_comparison}. Seed 5 failed to converge and is therefore excluded from the energy plots.

The results reveal significant deviations from the reference solution in both the elastic and fracture energies. This behaviour is attributed to the fact that the architecture of Manav \emph{et al.} was originally designed for second-order phase-field formulations. In contrast, the anisotropic fracture problems considered in the present work are governed by a fourth-order phase-field model, which demands considerably greater network expressivity to accurately resolve the higher-order differential operators involved.

The solutions for the network architecture proposed by Manav \emph{et al.} converged in $1256\pm374$ (mean $\pm$ std) minutes across the 10 random seeds.

\subsubsection{Crack propagation in layered orthotropic media}  \label{sec:sq_plate_layered}

Next, a layered plate composed of orthotropic materials with alternating orientations is examined. The objective of this example is to demonstrate that the proposed DRM framework can accurately solve the underlying optimisation problem even when the crack path undergoes sudden changes in direction. Such situations commonly occur in heterogeneous anisotropic media and may lead to crack kinking at material interfaces. For this example, the early stopping tolerance was tightened to $\Delta\mathcal{L}_{\mathrm{rel}} < 5\times10^{-6}$.

The plate geometry and boundary conditions remain identical to those described in Section~\ref{sec:sq_plate_tension}. The material domain is divided into three regions by vertical interfaces located at 1/3rd and 2/3rds of the plate. Each region follows an orthotropic material symmetry with an orientation of $20^{\circ}$, but with alternating preferred fracture directions across the interface, as shown in \FIG{fig:sq_plate_alternating}. This configuration introduces a discontinuity in the preferred fracture direction. As a result, the propagating crack experiences an abrupt change in the fracture energy landscape when crossing the interface.

The evolution of elastic and fracture energies is analysed as a function of the applied displacement. The predicted phase-field distributions obtained with the proposed DRM are compared with reference solutions from the FEM. For clarity, the results are presented using three subfigures: the energy evolution, the FEM phase-field distribution, and the DRM phase-field distribution.

\begin{figure}[htbp]
	\centering
    \begin{tabular}{>{\centering\arraybackslash}m{0.45\textwidth} 
            >{\centering\arraybackslash}m{0.25\textwidth} 
        >{\centering\arraybackslash}m{0.25\textwidth}}
		\subfloat[\label{fig:layered_case:a}]{
        \includegraphics[width=0.45\textwidth]{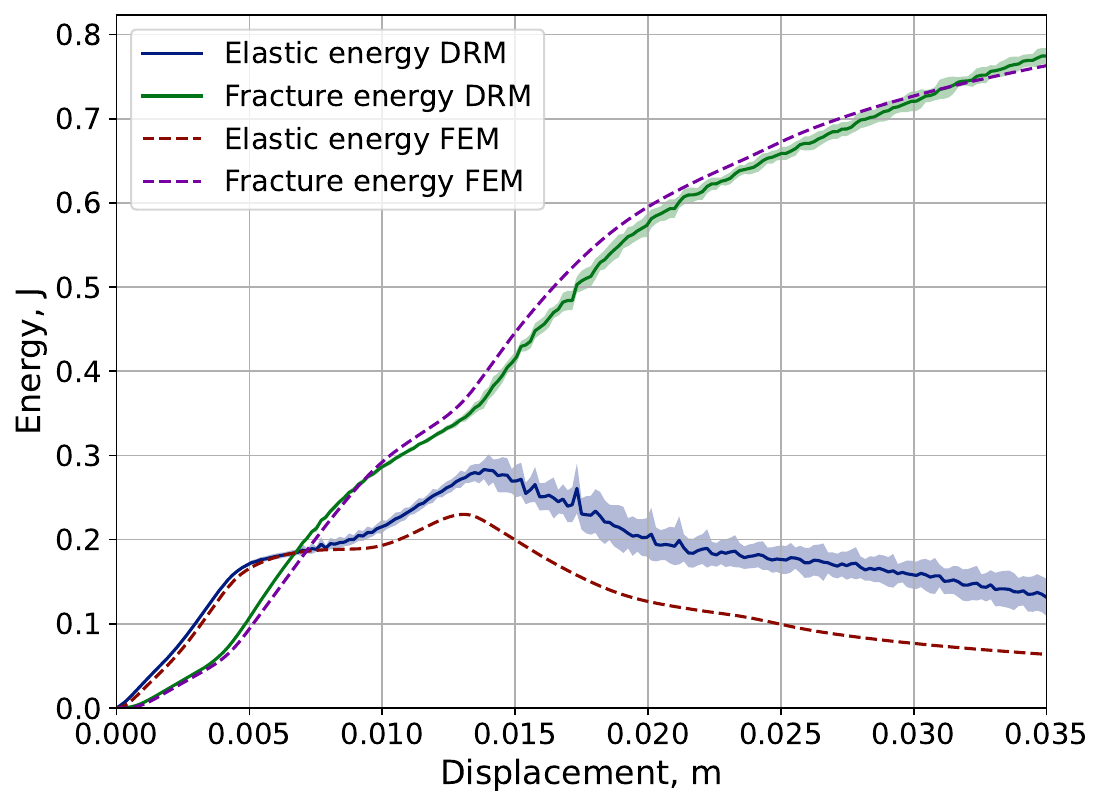}} &
        \subfloat[\label{fig:layered_case:b}]{
    \includegraphics[width=0.25\textwidth]{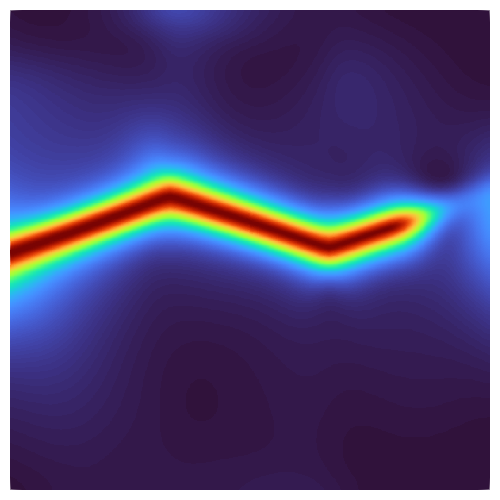}} &
        \subfloat[\label{fig:layered_case:c}]{
    \includegraphics[width=0.25\textwidth]{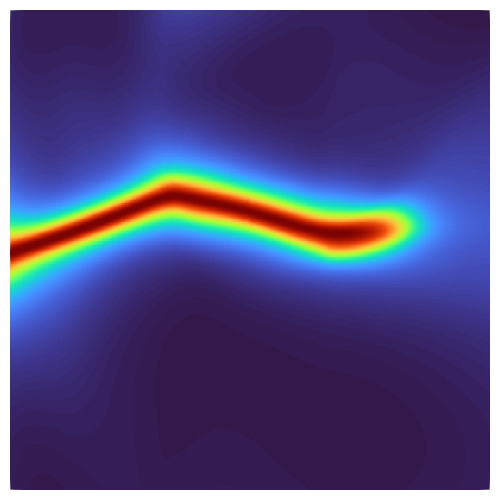}} \\
    \multicolumn{3}{c}{\includegraphics[width=0.25\textwidth]{colorbar.png}}
	\end{tabular}
	\caption[]{Results for the layered orthotropic material case. 
	\subref{fig:layered_case:a} Evolution of elastic and fracture energies as a function of displacement. 
	\subref{fig:layered_case:b} Phase-field distribution obtained using the FEM. 
\subref{fig:layered_case:c} Phase-field distribution obtained using the proposed DRM at a displacement of $0.035~\text{m}$. The fracture trajectory is generated by seed number $7$.}
    \label{fig:layered_case}
\end{figure}

The numerical results are presented in \FIG{fig:layered_case}. The energy evolution shown in \FIG{fig:layered_case:a} follows the typical behaviour of phase-field fracture simulations. The elastic energy increases with the applied displacement until crack initiation occurs. As the crack propagates through the layered medium, fracture energy increases while the elastic energy decreases due to stress redistribution.

The FEM phase-field distribution shown in \FIG{fig:layered_case:b} illustrates the interaction between the propagating crack and the material interface. When the crack reaches the interface, its trajectory changes abruptly due to the change in orthotropic orientation. This results in crack kinking and subsequent reorientation of the crack path. The resulting trajectory therefore reflects the underlying anisotropic material structure.

The DRM prediction shown in \FIG{fig:layered_case:c} reproduces this behaviour with excellent agreement up to the third material layer, beyond which the predicted crack width becomes excessively diffuse. We attribute this to coupling effects between the phase- and displacement fields; the use of smooth activation functions hinders the displacement field from resolving the discontinuity along the crack path, which in turn impedes convergence of the sharp turn in the phase-field, resulting in the fracture pattern observed in \FIG{fig:layered_case:c}.

Nevertheless, the general shape of the predicted crack trajectory remains accurate, capturing even the onset of the final crack deflection in the last layer. This demonstrates that the proposed neural framework is capable of navigating the complex optimisation landscape introduced by the material interfaces and the associated abrupt changes in crack path to a considerable extent.

The DRM solutions converged in $1605\pm119$ (mean $\pm$ std) minutes across the 10 random seeds.

\subsubsection{Convergence analysis} \label{sec:sq_plate_convergence}

\paragraph{Influence of displacement increment size} \label{sec:sq_plate_loading_steps}

\begin{figure}[htbp]
    \label{fig:loading_steps}
	\centering
    \begin{tabular}{>{\centering\arraybackslash}m{0.45\textwidth} 
            >{\centering\arraybackslash}m{0.45\textwidth}}
        \subfloat[\label{fig:disp_conv:elastic}]{
            \includegraphics[width=0.45\textwidth]{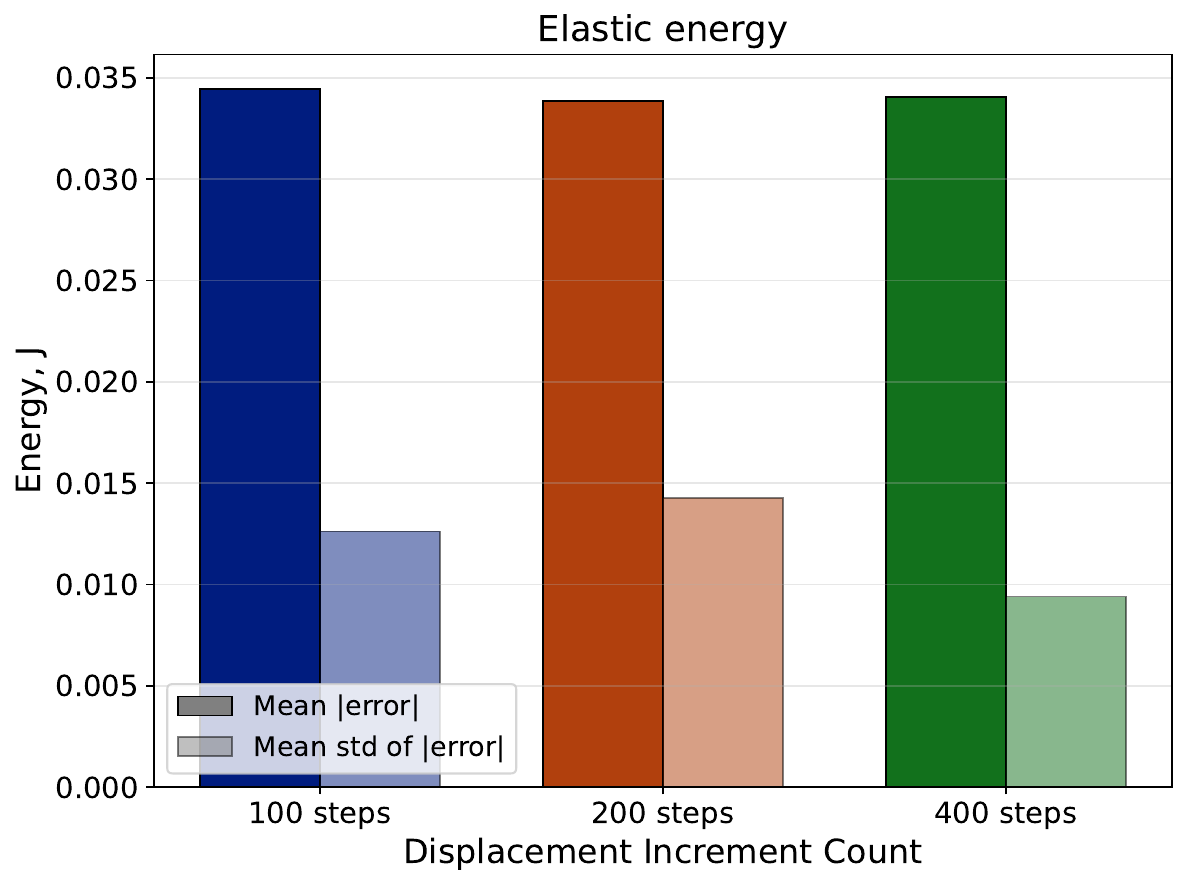}} &
        \subfloat[\label{fig:disp_conv:fracture}]{
            \includegraphics[width=0.45\textwidth]{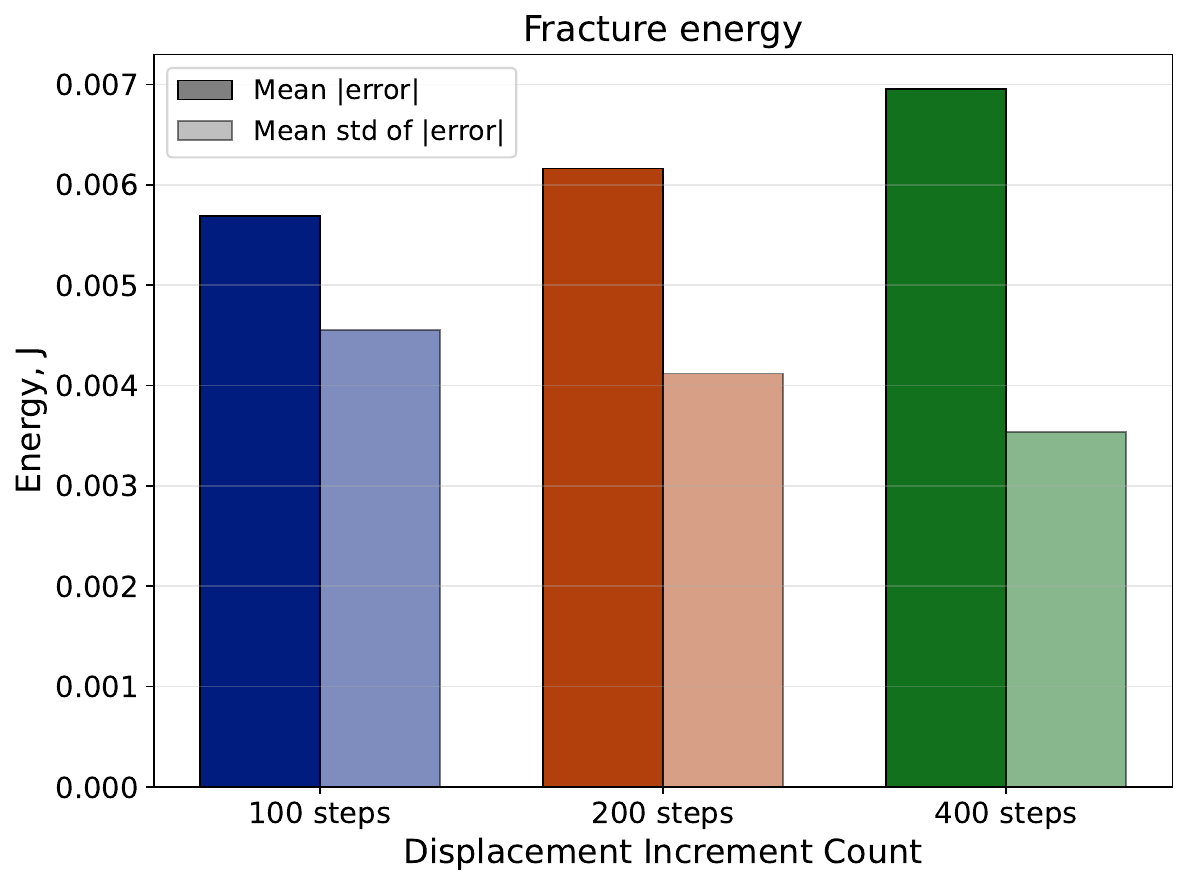}} \\
        \multicolumn{2}{c}{
            \begin{tabular}{>{\centering\arraybackslash}m{0.25\textwidth} 
                    >{\centering\arraybackslash}m{0.25\textwidth} 
                    >{\centering\arraybackslash}m{0.25\textwidth}}
                \subfloat[\label{fig:disp_conv:100}]{
                    \includegraphics[width=0.25\textwidth]{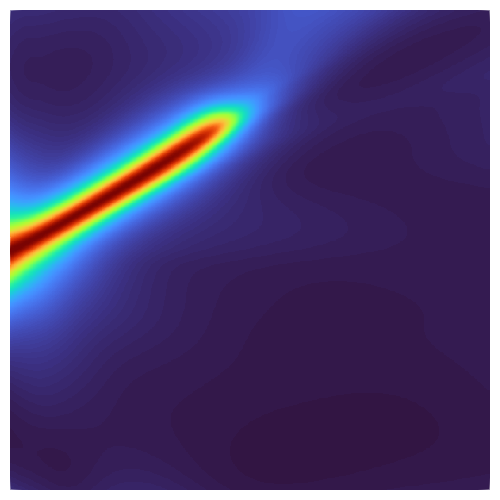}} &
                \subfloat[\label{fig:disp_conv:200}]{
                    \includegraphics[width=0.25\textwidth]{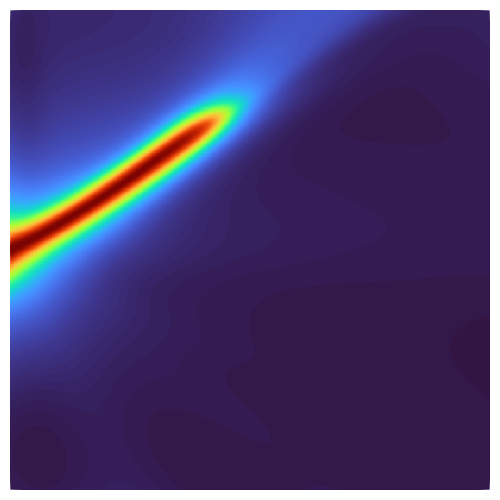}} &
                \subfloat[\label{fig:disp_conv:400}]{
                    \includegraphics[width=0.25\textwidth]{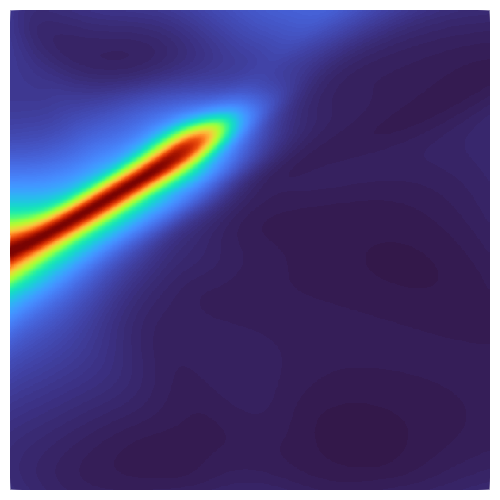}}
            \end{tabular}
        } \\
        \multicolumn{2}{c}{\includegraphics[width=0.25\textwidth]{colorbar.png}}
	\end{tabular}
	\caption[]{Displacement step convergence study. \subref{fig:disp_conv:elastic} Mean absolute error and standard deviation of elastic energy, and \subref{fig:disp_conv:fracture} fracture energy, evaluated over the whole simulation length. Phase-field distributions for \subref{fig:disp_conv:100} 100 displacement increments, \subref{fig:disp_conv:200} 200 increments, and \subref{fig:disp_conv:400} 400 increments. All fracture trajectories are generated by seed number 2.}
    \label{fig:disp_conv}
\end{figure}

The influence of displacement increment size on the predicted crack path and energy evolution is investigated by varying the number of increments used to reach the maximum applied displacement. The FEM reference solution uses 500 increments, while DRM predictions use $100$, $200$, and $400$ increments for the same mesh size of $1/201$ m. For each case, the absolute error between the DRM and FEM energy values is computed at every displacement step, and the mean and standard deviation across the ensemble are averaged over the full loading history. The results are presented in \FIG{fig:disp_conv}.
The mean error and standard deviation of the elastic energy are similar across all three cases. For the fracture energy, the mean error increases with the number of increments, while the standard deviation decreases. 

This trend indicates that finer temporal discretisation does not necessarily improve accuracy. The DRM solution at each increment does not always converge to a global minimum. As a result, errors may accumulate over the loading history. Increasing the number of increments allows such errors to accumulate further.

The phase-field distributions shown in \FIG{fig:disp_conv:100}--\FIG{fig:disp_conv:400} illustrate the effect of increment size on the predicted crack path. The fracture trajectories obtained in all cases are in close agreement. In terms of computational time, the DRM converged in $207\pm39$, $293\pm41$, and $424\pm43$ minutes (mean $\pm$ standard deviation over $10$ random seeds) for $100$, $200$, and $400$ displacement increments, respectively.

\paragraph{Influence of mesh discretisation}  \label{sec:sq_plate_mesh}

The effect of mesh discretisation on crack path prediction and energy evolution is examined. The FEM reference solution is obtained using a mesh with element spacing $1/201~\text{m}$, while the DRM predictions are generated using meshes with element spacing $1/101~\text{m}$, $1/201~\text{m}$, and $1/301~\text{m}$. For each case, the absolute error between the DRM and FEM energy values is computed at every displacement step, and the mean and standard deviation across the ensemble are averaged over the full loading history. 

\begin{figure}[htbp]
	\centering
    \begin{tabular}{>{\centering\arraybackslash}m{0.45\textwidth} 
            >{\centering\arraybackslash}m{0.45\textwidth}}
        \subfloat[\label{fig:mesh_conv:elastic}]{
            \includegraphics[width=0.45\textwidth]{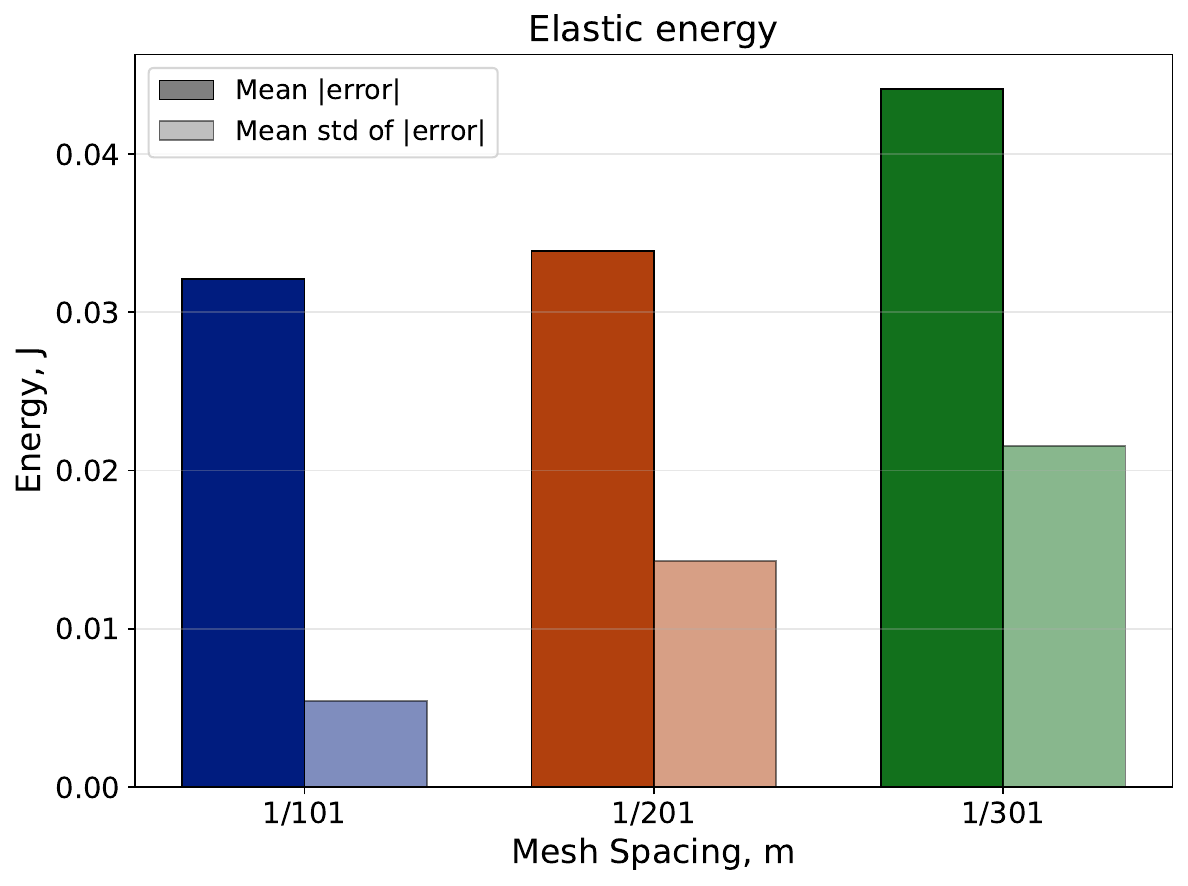}} &
        \subfloat[\label{fig:mesh_conv:fracture}]{
            \includegraphics[width=0.45\textwidth]{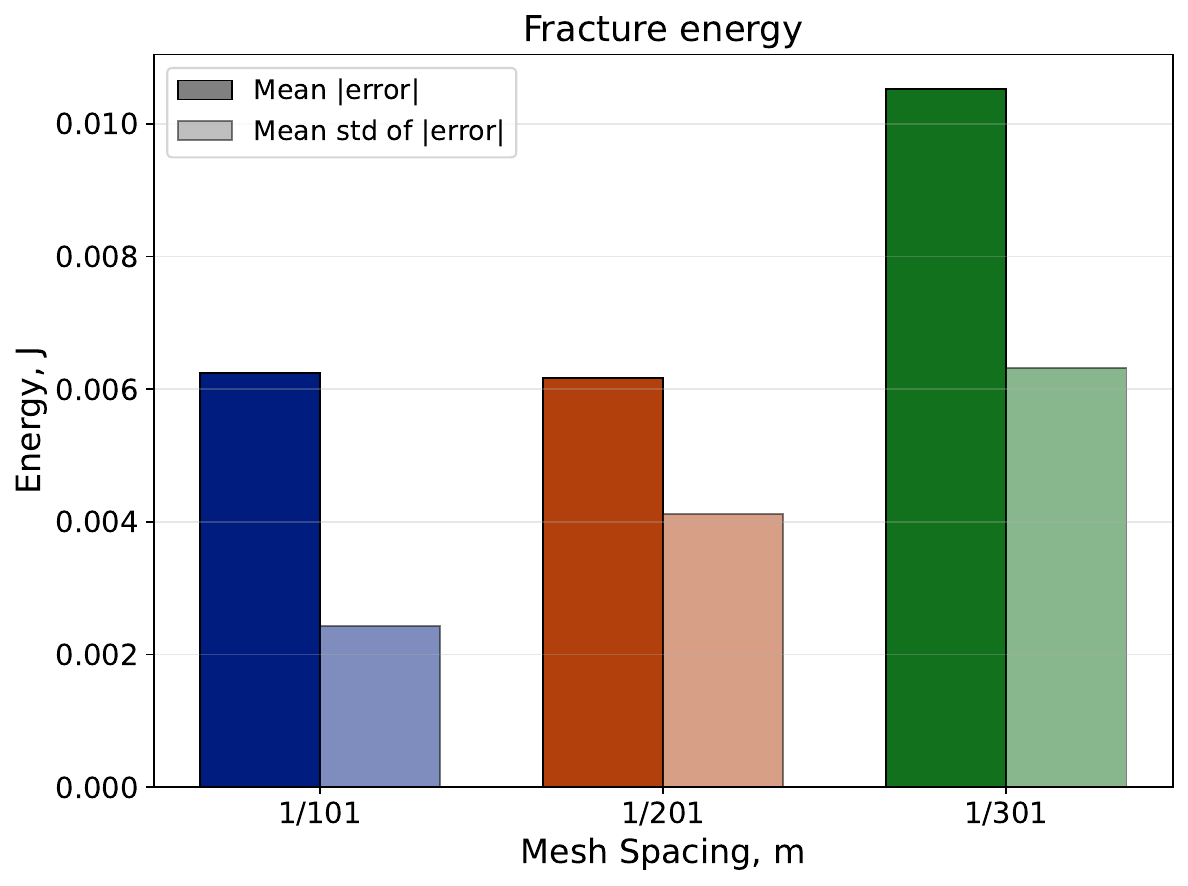}} \\
        \multicolumn{2}{c}{
            \begin{tabular}{>{\centering\arraybackslash}m{0.25\textwidth} 
                    >{\centering\arraybackslash}m{0.25\textwidth} 
                    >{\centering\arraybackslash}m{0.25\textwidth}}
                \subfloat[\label{fig:mesh_conv:101}]{
                    \includegraphics[width=0.25\textwidth]{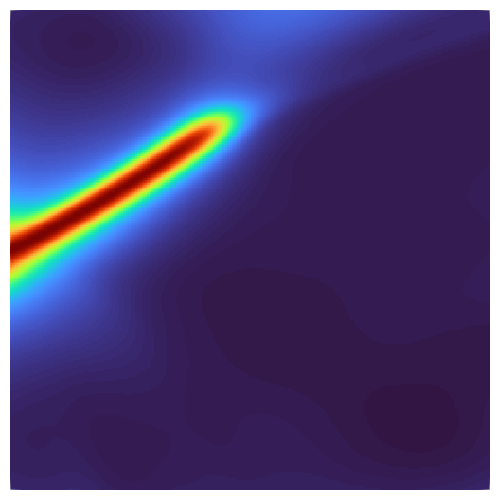}} &
                \subfloat[\label{fig:mesh_conv:200}]{
                    \includegraphics[width=0.25\textwidth]{convergence_reference.png}} &
                \subfloat[\label{fig:mesh_conv:301}]{
                    \includegraphics[width=0.25\textwidth]{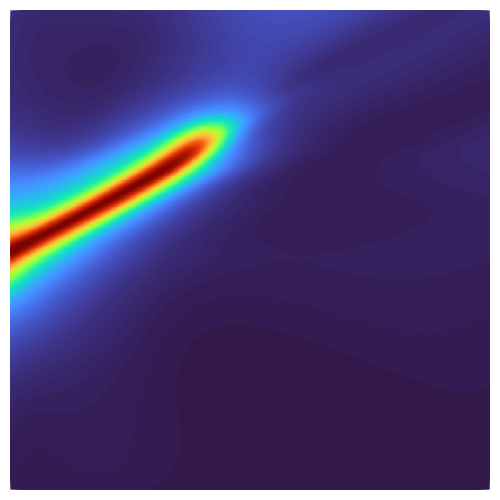}}
            \end{tabular}
        } \\
        \multicolumn{2}{c}{\includegraphics[width=0.25\textwidth]{colorbar.png}}
	\end{tabular}
	\caption[]{Mesh convergence study. \subref{fig:mesh_conv:elastic} Mean absolute error and standard deviation of elastic energy, and \subref{fig:mesh_conv:fracture} fracture energy, evaluated over the whole simulation length. Phase-field distributions for \subref{fig:mesh_conv:101} mesh with 101 nodes, \subref{fig:mesh_conv:200} 200 nodes, and \subref{fig:mesh_conv:301} 301 nodes. All fracture trajectories are generated by seed number 2.}
    \label{fig:mesh_conv}
\end{figure}

The results are presented in \FIG{fig:mesh_conv}.
There is little variation in the mean error of both elastic and fracture energies when refining from element spacing $1/101~\text{m}$ to $1/201~\text{m}$. A significant increase in both the mean absolute error and the standard deviation is observed when refining to $1/301$. The similarity between the $1/101~\text{m}$ and $1/201~\text{m}$ results is notable, as the convention in the phase-field fracture literature is to choose an element size of approximately $l_0/2$, which corresponds to the $1/201~\text{m}$ mesh. This suggests that the proposed DRM framework can produce comparable results with a significantly coarser mesh than is typically required in FEM formulations.

The standard deviation increases monotonically with mesh refinement. This behaviour is attributed to dilution of the network's representational capacity in the fracture region. Uniform mesh refinement increases the number of nodes quadratically, while the fracture process zone remains localised. Since the network inputs consist of standalone $x$- and $y$-coordinates, training is distributed across all nodes. The fracture region therefore constitutes a progressively smaller fraction of the domain.

The phase-field distributions shown in \FIG{fig:mesh_conv:101}--\FIG{fig:mesh_conv:301} illustrate the effect of mesh refinement on the predicted crack path. The fracture trajectories for element spacings of $1/101~\mathrm{m}$ and $1/201~\mathrm{m}$ are in close agreement with the reference solution. The $1/301~\mathrm{m}$ case shows increased variability in the crack region, with the trajectory developing at a slightly shallower angle. This behaviour suggests that excessive mesh refinement may increase sensitivity to local minima in the optimisation landscape.

The DRM converged in $74\pm6$, $293\pm41$, and $734\pm140$ minutes (mean $\pm$ standard deviation over 10 random seeds) for meshes with element spacings of $1/101~\mathrm{m}$, $1/201~\mathrm{m}$, and $1/301~\mathrm{m}$, respectively

\section{Conclusions} \label{sec:conclusions}

This work presented a variational physics-informed deep learning framework for the phase-field modelling of brittle fracture in anisotropic media. The proposed formulation extends the DRM to higher-order anisotropic phase-field models through a generalised crack density functional. A hybrid neural-spline strategy was introduced to enable stable evaluation of higher-order spatial derivatives while preserving the variational structure of the governing problem.

The methodology was assessed through numerical examples involving isotropic, cubic, and orthotropic fracture behaviour. The results demonstrated that the proposed framework captures crack nucleation and propagation under a range of anisotropic conditions. Direction-dependent fracture paths were obtained in good agreement with finite element reference solutions. In heterogeneous anisotropic media, the method reproduced crack kinking phenomena consistent with expected physical behaviour. Agreement with finite element solutions was observed for both phase-field distributions and energy evolution.

The study also highlighted several important numerical aspects of variational deep learning for fracture. The choice of optimisation strategy was found to influence stability and crack evolution under incremental loading, with first-order methods providing more consistent behaviour in non-convex settings. Temporal discretisation was not found to systematically improve accuracy, as errors may accumulate across increments when local minima are encountered. Similarly, excessive mesh refinement can reduce the effective resolution of the fracture process zone within the network approximation, leading to increased variability in the predicted crack path.

Future work will focus on extensions to three-dimensional fracture problems and more advanced material models. Further developments will consider adaptive sampling strategies, improved optimisation schemes for strongly non-convex energy landscapes, and the integration of experimental data within hybrid physics-informed learning frameworks.

\section*{CRediT authorship contribution statement}
\textbf{N. Plungė}: Conceptualisation, Data curation, Formal analysis, Investigation, Methodology, Software, Validation, Visualization, Writing - original draft.  \textbf{P. Brommer}: Conceptualisation, Methodology, Supervision, Writing - review \& editing. \textbf{R. S. Edwards}: Conceptualisation, Methodology, Supervision, Writing - review \& editing. \textbf{E. G. Kakouris}: Conceptualisation, Funding acquisition, Methodology, Supervision, Writing - review \& editing.

\section*{Declaration of competing interest}
The authors declare that they have no known competing financial interests or personal relationships that could have influenced the work reported in this paper.

\section*{Data Availability Statement}
The input data used in this study and the code used to generate the FEM and DRM results are available through the associated Zenodo repository.\cite{plunge_2026_19553764} Output data are presented in the figures and can be provided upon reasonable request.

\section*{Acknowledgements}
Computing facilities were provided by the Scientific Computing Research Technology Platform of the University of Warwick. GPU calculations were performed using the Sulis Tier 2 HPC platform hosted by the Scientific Computing Research Technology Platform at the University of Warwick. Sulis is funded by EPSRC Grant EP/T022108/1 and the HPC Midlands+ consortium. The authors also acknowledge the support of the Engineering and Physical Sciences Research Council (EPSRC) through the HetSys Centre for Doctoral Training in Modelling of Heterogeneous Systems (EP/S022848/1), project reference 2886009.

\bibliography{mybibfile.bib}   
\bibliographystyle{unsrt}

\appendix
\section{Constitutive matrices for anisotropic elasticity}\label{app:constitutive}

The constitutive matrices are expressed in Voigt notation and are given in the material principal axes. The in-plane formulation corresponds to a two-dimensional setting under small-strain kinematics.

\paragraph{Cubic material}
For the cubic material, the in-plane constitutive matrix is written as
\begin{equation}
\mathbf{D}^{\mathrm{el}}_{\mathrm{cubic}} =
\begin{bmatrix}
\dfrac{E(1-\nu)}{(1+\nu)(1-2\nu)} & \dfrac{E\nu}{(1+\nu)(1-2\nu)} & 0 \\[8pt]
\dfrac{E\nu}{(1+\nu)(1-2\nu)} & \dfrac{E(1-\nu)}{(1+\nu)(1-2\nu)} & 0 \\[8pt]
0 & 0 & G
\end{bmatrix},
\end{equation}
where $E$ and $\nu$ denote the Young’s modulus and Poisson’s ratio, respectively, and $G$ is the in-plane shear modulus. The material reduces to the isotropic case when $G = E/[2(1+\nu)]$. Otherwise, prescribing $G$ independently introduces cubic anisotropy in the in-plane response.

\paragraph{Orthotropic material}
For the orthotropic material, the in-plane constitutive matrix is expressed as
\begin{equation}
\mathbf{D}^{\mathrm{el}}_{\mathrm{orth}} =
\begin{bmatrix}
C_{11} & C_{12} & 0 \\
C_{12} & C_{22} & 0 \\
0 & 0 & G_{12}
\end{bmatrix},
\end{equation}
where the stiffness coefficients are obtained from the corresponding three-dimensional orthotropic elasticity tensor as
\begin{equation}
C_{11}=\frac{1-\nu_{23}\nu_{32}}{E_{22}E_{33}\Delta}, \qquad
C_{22}=\frac{1-\nu_{13}\nu_{31}}{E_{11}E_{33}\Delta}, \qquad
C_{12}=\frac{\nu_{21}+\nu_{31}\nu_{23}}{E_{22}E_{33}\Delta},
\end{equation}
and
\begin{equation}
\Delta=
\frac{
1-\nu_{12}\nu_{21}-\nu_{23}\nu_{32}-\nu_{13}\nu_{31}
-2\nu_{21}\nu_{32}\nu_{13}
}{E_{11}E_{22}E_{33}}.
\end{equation}

Here, $E_{ii}$ denote the Young’s moduli along the principal material directions, $\nu_{ij}$ are the Poisson ratios, and $G_{12}$ is the in-plane shear modulus. The reciprocal relations $\nu_{ji} = \nu_{ij} E_{jj}/E_{ii}$ are assumed to ensure symmetry of the constitutive tensor.

\end{document}